\documentclass[twocolumn, trackchanges]{aastex63}
\usepackage{cancel}                                                     %This allows commands to put lines through terms.
\usepackage{mathtools}                                                  %This adds a series of packages designed to enhance the appearance of documents containing a lot of mathematics.
\usepackage{amsmath, amssymb, amsfonts, amsthm}                         %This adds many math commands.
\usepackage{tikz}                                                       %This adds drawing features.
\usepackage{verbatim}                                                   %This adds the comment environment and reimplements the verbatim command.
\usepackage{ulem}                                                       %This adds sout.
\usepackage{float}                                                      %This adds full width footnotes in double column formats.
\usepackage{wrapfig}                                                    %This allows for text wrapping around figures.
\usepackage[utf8]{inputenc}                                             %This lets me type naïve.
\usepackage{enumitem}
\usepackage{graphicx}\graphicspath{{"figures"/}}                        %This allows the import and display of figures. %}\]
\renewcommand{\t}{\text}                                                %This is a shortcut for the text environment.
\renewcommand{\c}{\cdot}                                                %This is a shortcut for the dot operator.
\renewcommand{\P}{\mathbb{P}}                                           %This is a shortcut for blackboard P.
\renewcommand{\l}{\ell}                                                 %This is a shortcut for script l.
\renewcommand{\i}{\overline{i}}                                         %This is a shortcut for i conjugate.
\renewcommand{\j}{\overline{j}}                                         %This is a shortcut for j conjugate.
\renewcommand{\k}{\overline{k}}                                         %This is a shortcut for k conjugate.
\renewcommand{\l}{\overline{l}}                                         %This is a shortcut for k conjugate.
\renewcommand{\r}{\scriptr}                                             %This is a shortcut for script r.%}\]
\newcommand{\commentspace}[1]{}                                         %This makes a comment inline.
\newcommand{\be}{\begin{eqnarray}}                                      %This is a shortcut for beginning the equation array environment.
\newcommand{\ee}{\end{eqnarray}}                                        %This is a shortcut for ending the equation array environment.
\newcommand{\p}[1]{\left(#1\right)}                                     %This is a shortcut for parenthesis.
\newcommand{\pd}[1]{(#1)}                                               %This is a shortcut for normal parenthesis.
\newcommand{\set}[1]{\left\{#1\right\}}                                 %This is a shortcut for set notation.
\newcommand{\td}{\t{d}}                                                 %This is a shortcut for straight d.
\newcommand{\R}{\mathbb{R}}                                             %This is a shortcut for blackboard R.
\newcommand{\Z}{\mathbb{Z}}                                             %This is a shortcut for blackboard Z.
\newcommand{\N}{\mathbb{N}}                                             %This is a shortcut for blackboard N.
\newcommand{\C}{\mathbb{C}}                                             %This is a shortcut for blackboard C.
\newcommand{\Q}{\mathbb{Q}}                                             %This is a shortcut for blackboard C.
\newcommand{\F}{\mathbb{F}}                                             %This is a shortcut for blackboard F.
\newcommand{\m}[1]{\begin{bmatrix*}[r] #1\end{bmatrix*}}                %This is a shortcut for a matrix.
\let\oldsqrt\sqrt                                                       %This redefines the sqrt function as oldsqrt.
\def\sqrt{\mathpalette\DHLhksqrt}                                       %This line and the ones following redefine the sqrt function to be prettier.
\def\DHLhksqrt#1#2{\setbox0=\hbox{$#1\oldsqrt{#2\,}$}%
    \dimen0=\ht0\advance\dimen0-0.2\ht0\setbox2%
    =\hbox{\vrule height\ht0 depth -\dimen0}{\box0\lower0.4pt\box2}}
\def\red#1 {\textcolor{red}{#1}\ }                                      %This makes text red.
\def\redcap#1 {\textcolor{red}{\uppercase{#1}}\ }                       %This makes text red and uppercase.
\shorttitle{Chaos in Three-Planet Systems}
\shortauthors{Rath et al.}
\begin{document}
% Title Page %\[{
\title{The Criterion for Chaos in Three-Planet Systems}
% Authors \[{
\author[0000-0003-4496-5958]{Jeremy Rath}
\affiliation{Dept. of Physics and Astronomy, Northwestern University, 2145 Sheridan Rd., Evanston, IL 60208, USA}
\affiliation{Center for Interdisciplinary Exploration and Research in Astrophysics (CIERA), 1800 Sherman, Evanston, IL 60201, USA}
\correspondingauthor{Jeremy~Rath}
\email{jeremyrath2020@u.northwestern.com}
\author[0000-0002-1032-0783]{Sam Hadden}
\affiliation{Canadian Institute for Theoretical Astrophysics, 60 St George St Toronto, ON M5S 3H8, Canada}
\affiliation{Center for Astrophysics $\vert$ Harvard \& Smithsonian, Cambridge, MA 02138, USA}
\author[0000-0003-4450-0528]{Yoram Lithwick}
\affiliation{Dept. of Physics and Astronomy, Northwestern University, 2145 Sheridan Rd., Evanston, IL 60208, USA}
\affiliation{Center for Interdisciplinary Exploration and Research in Astrophysics (CIERA), 1800 Sherman, Evanston, IL 60201, USA}
% }\]

\begin{abstract}
We establish the criterion for chaos in three-planet systems,
    for systems similar to those discovered by the Kepler spacecraft.
Our main results are as follows:
    (i) The simplest criterion, which is based on overlapping mean motion resonances (MMR's),
        only agrees with numerical simulations at a very crude level.
   (ii) Much greater accuracy is attained by considering neighboring MMR's that do not overlap.
        We work out the width of the chaotic zones around each of the neighbors, and also provide simple approximate expressions for the widths.
  (iii) Even greater accuracy is provided by the overlap of three-body resonances (3BR's),
        which accounts for  fine-grained structure seen in maps from N-body simulations,
        and also predicts the Lyapunov times.
Previous studies conflict on whether overlap of MMR's or of 3BR's drive interplanetary chaos.
We show that both do, and in fact they are merely different ways of looking at the same effect.
(iv) We compare both criteria with high-resolution maps of chaos from N-body simulations, and show that they agree at a high level of detail.
\end{abstract}

% }\]

\section{Introduction}\label{sec:intro}
% Examples of Chaotic Systems \[{
In the Solar System, the planets' orbits are chaotic \citep{Laskar1996, LaskarGastineau2009, HolmanMurray1996}.
Many exoplanetary systems are also chaotic, or at least close to the threshold for chaos, e.g., Kepler-11 \citep{MahajanWu2014, Yee2021}, Kepler-36 \citep{Deck2012}, Kepler-102 \citep{VolkMalhotra2020}, GJ876 \citep{Gozdziewski2002, Batygin2015b}, and Kepler-24, -85, and -444 \citep{Yee2021}.
The chaos cannot be too violent, because otherwise the systems would  not survive for billions of years.
That raises the question of how planets were emplaced into such delicate configurations.
It also opens the door to using observed planetary configurations to learn about their early history. 
But our understanding of how chaos works in planetary systems is  not yet sufficient to answer such questions.
% }\]

% Overlap Criterion \[{
A simple and powerful criterion for the presence of chaos is {\it resonance overlap} \citep{WalkerFord1969, 1979_chirikov}.
Resonance overlap accounts for much of the chaos in planetary systems. But which resonances overlap depends on the system,
    as planetary systems harbor a variety of different kinds of resonances.
In nearly circular two-planet systems, decreasing the planets' separation drives them toward chaos. 
There is a critical separation inside of which the orbits are always chaotic,
    and its value is determined by where neighboring first-order mean motion resonances (MMR's) overlap \citep{1980_wisdom,2013_deck}.
Two-planet systems can also be driven towards chaos by increasing their eccentricity, 
    because the widths of MMR's increase with increasing eccentricity. 
The critical eccentricity for chaos has been determined by, e.g., \cite{Mardling2008, 2018_hadden}.
It depends not only on first-order MMR's, but on higher-order MMR's as well.
In systems of more than two planets, \cite{Tamayo2021} suggest that chaos is generally attributable to the overlap of two-body MMR's---similar to the two-planet case.
% }\]

% Figure: Lots of Crossings - Numerical \[{
\begin{figure}[!t]
    \includegraphics[width=0.99\columnwidth]{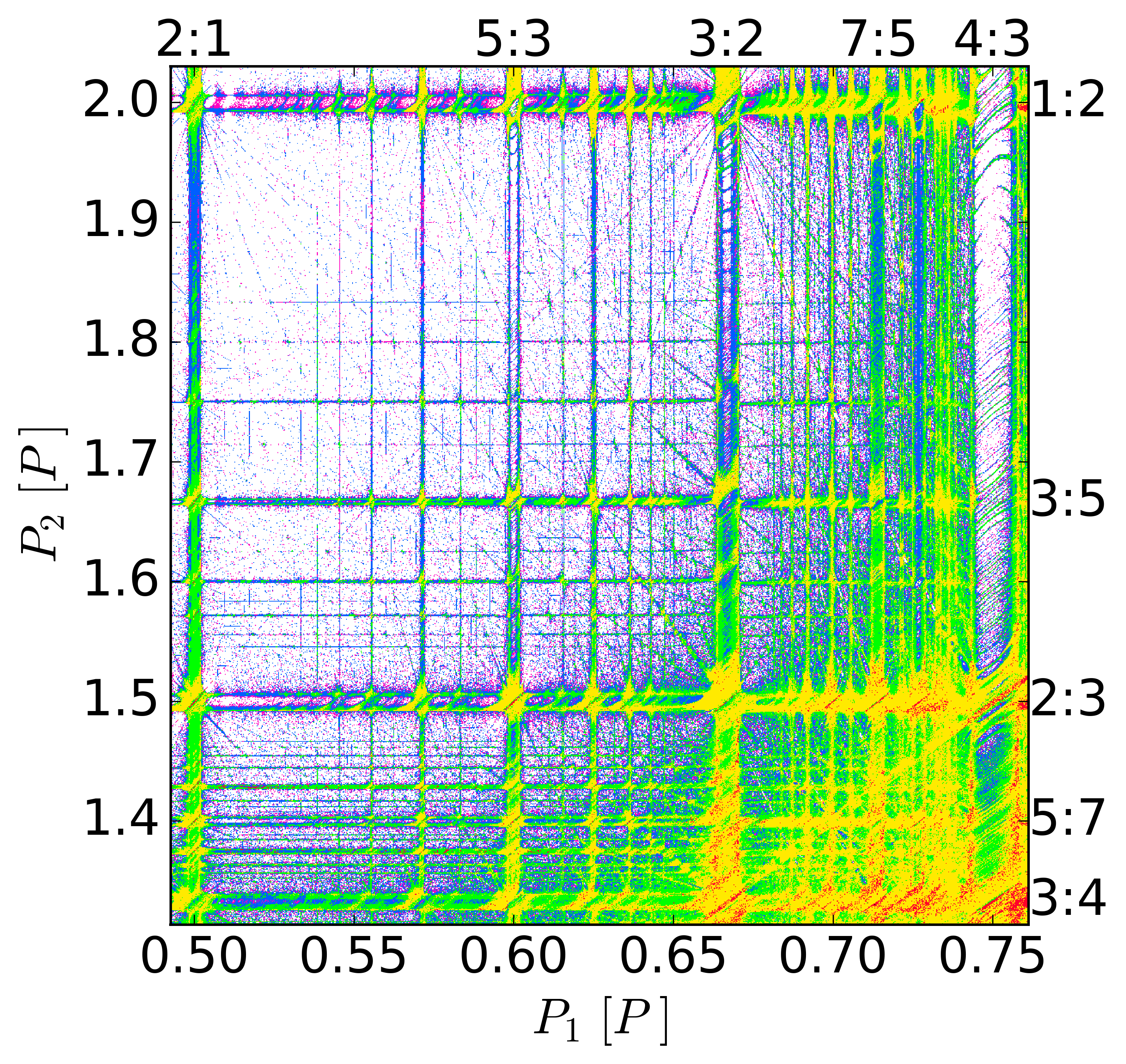}
    \vspace{-1 pt}
    \includegraphics[width=0.99\columnwidth]{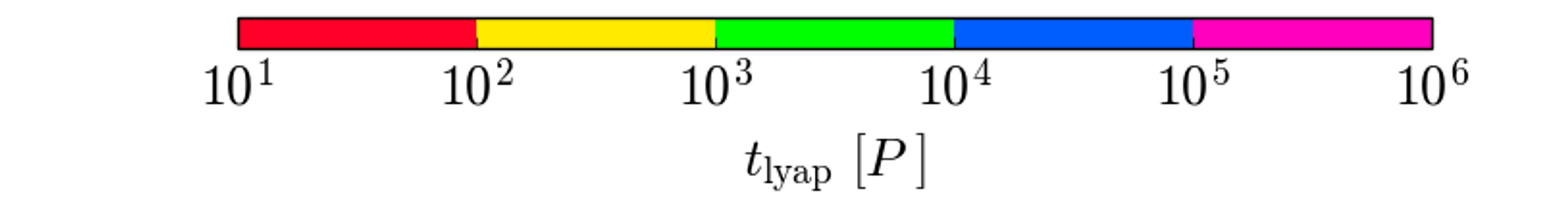}
    \centering
    \caption{Lyapunov times for our fiducial three-planet system in the PP-plane: 
             The axes give the initial orbital periods of the inner and outer planets,
                in units of the middle planet's period, and the Lyapunov time is as colored.
             The locations of 1st and 2nd order MMR's are on the top and right axes.
             Broadly speaking, MMR's tend to be chaotic, and overlapping resonances enhance chaos.}    \label{kepler-11}
\end{figure}
% }\]

% Other Resonances \[{
Overlap of other kinds of resonances also produces chaos.
In  the inner Solar System, chaos is caused by the overlap of secular resonances \citep{Laskar1996, LithwickWu2011, Batygin2015}.
And in the outer Solar System, chaos has been attributed to the overlap of three-body resonances \citep[3BR's;][]{1999_murray_holman}.
\cite{2011_quillen} and \cite{Petit2020} suggest that chaos in generic planetary systems is driven by overlapping 3BR's,
    in contrast to the claim of \cite{Tamayo2021} that overlapping two-body MMR's are responsible.
One of the goals of this paper is to resolve this apparent discrepancy.
% }\]

% Discuss Established Theory \[{
Chaos in generic Hamiltonian systems is understood at a much deeper level than simply resonance overlap
\citep[see, e.g., textbooks and reviews of][]{1979_chirikov, 1983_lichtenberg, 1985_escande, 2007_zaslavsky}.
Two concepts in particular play a prominent role in this paper.
First, nearby resonances that do not overlap excite chaos in the vicinity of their separatrices.
The extent of the chaotic zone around each separatrix is readily calculable by applying the ``whisker'' or ``separatrix'' map (which we shall also refer to as the ``theory of kicks'').
One might suspect that the resulting correction to the resonance overlap criterion is at best an order-unity one.
But, as we shall demonstrate, resonance overlap is only approximately correct when the two resonances have comparable strengths.
In the more general case that one resonance is much stronger than the other,
   the correction becomes very large. 
The second concept is that of secondary resonances:
    two nearby resonances excite secondary resonances,
    and those secondaries can overlap and lead to chaos.\footnote{
        In a  planetary system, the secondary resonances from two nearby MMR's are 3BR's.}
There are then two predictions for the extent of the chaotic zone---from the theory of kicks and from secondary resonances---and they have been shown to agree with each other \citep[e.g.][]{1983_lichtenberg}.
While these concepts are detailed in standard references,
    we provide a self-contained discussion in this paper in order to aid the non-expert.
We also develop some improvements on the standard theory, as summarized in \S \ref{sec:recap}.
% }\]

% Figure: Lots of Crossings - Overlap \[{
\begin{figure}[!t]
    \includegraphics[width=0.99\columnwidth]{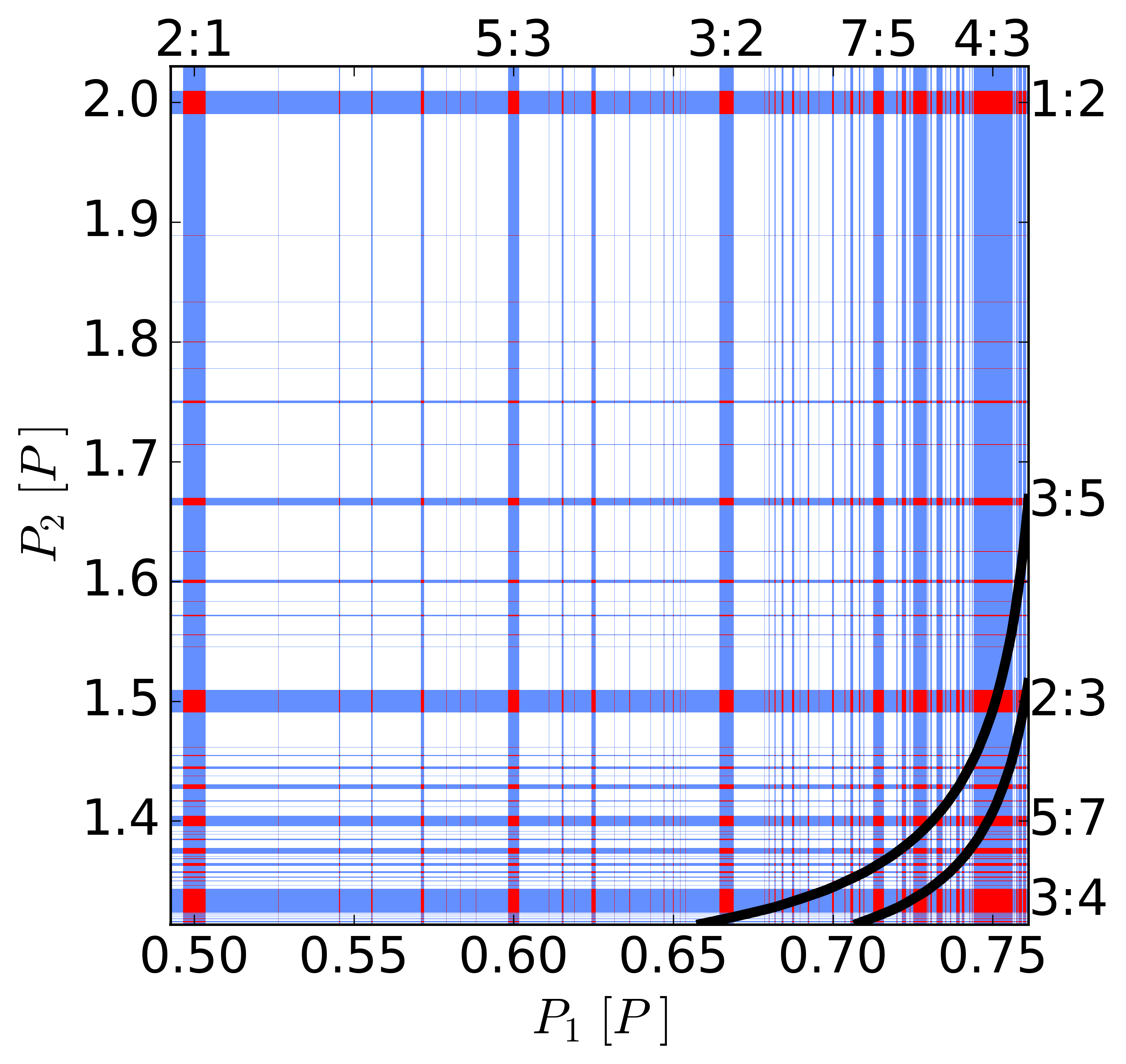}
    \centering
    \caption[]{Locations and widths of two-body MMR's for the system depicted in Fig. \ref{kepler-11}.
             The vertical blue bands are resonances between the inner two planets;
                the horizontal bands are between the outer two;
                and overlapping zones are red. The first- and second-order MMR's are labeled;
                    but all resonances are shown in blue, unless their width is less than the pixel size.
             \cite{Tamayo2021} provide a semianalytic formula for the chaos criterion based on two-body MMR overlap.
    Their result is shown by the two black curves in the lower-right corner.
             The lower curve is based on the initial eccentricities, and the upper is based on the maximum eccentricities reached via purely secular evolution.
             The curves were made with D. Tamayo's code {SPOCK}, publicly available at \href{https://github.com/dtamayo/spock}{github.com/dtamayo/spock}.}
    \label{kepler-11ro}
\end{figure}
% }\]

% Paper Outline \[{
The outline of this paper is as follows.
In \S \ref{sec:ooc}, we use numerical integrations to map out the chaos of a fiducial three-planet system.
We use that map as a touchstone for the theory of chaos developed in later sections.
In \S \ref{sec:simple}, we show how the dynamics near a crossing of two MMR's can be reduced to a much simpler model system: the perturbed pendulum.
In \S \ref{sec:theory}, we review the theory of kicks for the perturbed pendulum,
    and develop an ``improved'' theory that is applicable further away from the separatrix.
We then use that theory to predict the chaos seen in the touchstone map.
In \S \ref{sec:so}, we apply the theory of overlapping secondary resonances to predict chaos.
In the final section (\S \ref{sec:dis}), we provide a summary, assess the validity of our main approximations, compare to prior work, and discuss the outer Solar System.
% }\]

\section{Map of Chaos in a fiducial Three-Planet System}\label{sec:ooc}
% Discuss System Parameters \[{
We explore the chaotic behavior of a fiducial three-planet system.
The system we choose is similar to planets c, d, and e of Kepler-11,
    as given by the all-eccentric fit of Table S4 in \cite{2011_lissauer},
    with two modifications: the middle planet's mass is set to zero, and its eccentricity is boosted.
    (\citeauthor{2013_lissauer}  \citeyear{2013_lissauer} provide updated parameters for the Kepler-11 planets,
    but we do not use those.).
To be explicit, the masses of the three planets
are $\{ 12, 0, 9.3\}M_\oplus$;
    the eccentricities are $\{0.003, 0.070, 0.016\}$; 
    the angular orbital elements and stellar mass are set to those in \cite{2011_lissauer};
    and inclinations are set to zero.
Converting the middle planet to a test particle allows us to focus on its dynamics,
    while neglecting its backreaction on the other bodies.
As discussed in \S \ref{sec:va}, generalizing to a massive middle planet is straightforward.
% }\]

% Describe Figure 1 \[{
Fig. \ref{kepler-11} shows the result from a suite of integrations of the fiducial system in the period-period plane (``PP-plane''), i.e. the axes represent
    the initial orbital periods of the inner and outer planets ($P_1$ and $P_2$, respectively).
For comparison, the nominal orbital periods from \cite{2011_lissauer} are $P_1/P=0.57$ and $P_2/P=1.41$,
    where $P$ is the period of the middle planet.
Each point in Fig. \ref{kepler-11} represents an integration with given initial periods,
    performed with the WHFast integrator \citep{2015_rein} in REBOUND \citep{2012_rein}.
The color represents the Lyapunov time,
    which is determined by fitting the MEGNO with a least-squares fit \citep{2003_cincotta}.
In general, the Lyapunov time can vary in time,
    whereas for Fig. \ref{kepler-11} we wish to show the Lyapunov time towards the beginning of the simulation.
Therefore, we halt a simulation when the runtime exceeds ten times the Lyapunov time,
    which we have found is typically long enough to provide a sufficiently accurate measurement of the Lyapunov time. 
Systems in which the inferred Lyapunov time exceeds $10^6P$ are colored white in Fig. \ref{kepler-11};
    most of those are likely non-chaotic. 
% }\]

% Describe Goal of Paper \[{
The bulk of this paper is aimed at predicting the structure in Fig. \ref{kepler-11}.
We observe that much of the chaos is associated with MMR's.
For comparison, Fig. \ref{kepler-11ro} shows in blue the locations and widths of MMR's between the inner two planets (vertical bands) and between the outer two planets (horizontal bands),
    where the widths are taken from Eqs. (\ref{dPin}) and (\ref{dPout}) below.
The simplest predictor for chaos is resonance overlap \citep{1979_chirikov}. 
Fig. \ref{kepler-11ro} depicts the overlapping regions in red.
We observe that resonance overlap provides a rough guide to where the chaotic zones are in Fig. \ref{kepler-11}.
Resonance overlap is seen to be most successful when the widths of the two overlapping resonances are comparable.
But it does not explain, e.g., why many of the resonances in Fig. \ref{kepler-11} are chaotic even when there appears to be no nearby overlapping resonances; nor why there are diagonal chaotic bands in Fig. \ref{kepler-11}.
We shall address these issues, and others, below.
% }\]

\section{Chaos of a Perturbed Pendulum}\label{sec:simple}
\subsection{Reduction to Perturbed Pendulum}\label{sec:rpp}
% Introduce Toy Model \[{
In order to explain the structure seen in period-period maps (such as Fig. \ref{kepler-11}),
    we examine how a test particle behaves when it is affected by two MMR's,
    one with an interior planet (to be labeled by subscript 1) and one with an exterior planet (labeled 2).
As we show in Appendix \ref{sec:rh}, the test particle's Hamiltonian can be written as a {\it perturbed pendulum}:\footnote{
    The main approximation used to derive Eq. (\ref{toy_model}) is the ``pendulum approximation,'' whereby the coefficients of the cosine terms are assumed to be constant \citep{1999_murray_dermott}.
      We assess its validity in \S \ref{sec:va}.
      We also adopt a novel approximation, following \cite{2019_hadden}, that allows us to combine together all of the $k_1+1$ cosine terms associated with the inner resonance, and similarly all of the $k_2+1$ terms with the outer.
      Without this approximation, the form of Eq. (\ref{toy_model}) would only be applicable to the case when the inner and outer planets are circular, and the middle planet is eccentric.
      But with it, we may consider three eccentric planets as a nearly trivial extension of the circular case.
      See \S \ref{sec:ecce} for further details. \label{pendapprox}} 
    \begin{align}
    H(\phi, p, t) &= {\frac{p^2}{2} - {\epsilon_1}\cos(\phi)} - {\epsilon_2}\cos(r(\phi - \nu t)) \ ,
    \label{toy_model}
    \end{align}
    where the first two terms describe a simple pendulum,
    and the third is the perturbation.
Before studying the dynamics of the perturbed pendulum,
    we summarize how it is related to the orbital system under consideration:
% }\]

% Introduce Variables - Itemized \[{
\begin{itemize}
    \item The first cosine term is from the MMR with the inner planet, and the second is from the one with the outer planet. 
          The dimensionless coefficients ($\epsilon_1$ and $\epsilon_2$) are proportional to the corresponding planet's mass, and are both $\ll 1$.
    \item The parameter $\nu$ is proportional to the difference between (i) the test particle's mean motion if it is at nominal resonance with planet 1; and (ii) the corresponding quantity with planet 2 (Eq. \ref{eq:nudef}).
          The parameter $r$ is  a  ratio of integers: if we label the frequency ratio of the inner resonance by $j_1$:$j_1 - k_1$, and the one with the outer is $j_2$:$j_2 + k_2$ (for positive integer $j$'s and $k$'s), then $r=j_2/j_1$. 
    \item The canonical co-ordinate $\phi$ is the resonant argument of the inner planet's resonance (Eq. \ref{eq:defphi}),
              and the momentum $p$ is proportional to the difference between the test particle's semi-major axis and that at nominal resonance with the inner planet (Eq. \ref{eq:deflast}).
The argument of the second cosine,
    \begin{equation}
    \psi  \equiv r(\phi-\nu t)
    \end{equation}
    is the resonant argument of the outer planet's resonance.
\end{itemize}
% }\]

% Invert Hamiltonian \[{
The form of Eq. (\ref{toy_model}), with the first two terms referring to the inner planet's resonance,
is convenient when the inner planet dominates the dynamics, and the outer planet is treated as a perturbation.
But that is merely a convention.
By symmetry, one could swap $\epsilon_1$ and $\epsilon_2$ in Eq. (\ref{toy_model}) without affecting the dynamics, provided $\phi$, $p$, $r$, and $\nu$ are redefined by swapping the 1 and 2 indices in their definitions (Eqs. \ref{eq:rdef}--\ref{eq:deflast}).
Nonetheless, throughout this paper we retain the convention of Eq. (\ref{toy_model}),
    with $p$ and $\phi$ corresponding to the inner planet's resonance (the ``$\epsilon_1$-resonance'')
    and $\psi$ to the outer planet's (the ``$\epsilon_2$-resonance'').

% }\]

% Reduce to 3 Parameters \[{
Equation (\ref{toy_model}) has four parameters ($\epsilon_1$, $\epsilon_2$, $r$, and $\nu$).
That may be reduced to three by rescaling  $p$, $H$, $t$, and $\nu$ to reduce Eq. (\ref{toy_model}) to $H(\phi, p, t) = {p^2\over 2} - \cos(\phi) - {\epsilon_2 \over \epsilon_1} \cos(r(\phi - \nu t))$. 
Therefore in analyzing Eq. (\ref{toy_model}), one may set $\epsilon_1\rightarrow 1$ without loss of generality.
Despite that, we shall retain $\epsilon_1$ for most of the paper,
    with the principal exceptions being \S \ref{sec:ckc} and Appendices \ref{sec:MA}--\ref{sec:srd},
    where we set $\epsilon_1=1$ to reduce clutter.\footnote{
    If one analyzes a perturbed pendulum with $\epsilon_1=1$,
    the results are easily generalized to arbitrary values of $\epsilon_1$ by setting
    \begin{eqnarray*}
    \set{ \epsilon_2,p,\nu,t,H}\rightarrow 
    \set{ {\epsilon_2\over\epsilon_1},{p\over\sqrt{\epsilon_1}},
     {\nu\over \sqrt{\epsilon_1}},
     t\sqrt{\epsilon_1},
     {H\over \epsilon_1}} \ .
     \end{eqnarray*}
     \label{ftn:cov}}
% \]}

\subsection{Surfaces of Section}\label{sec:numst}
\begin{figure}[!t]
    \includegraphics[width=0.95\columnwidth]{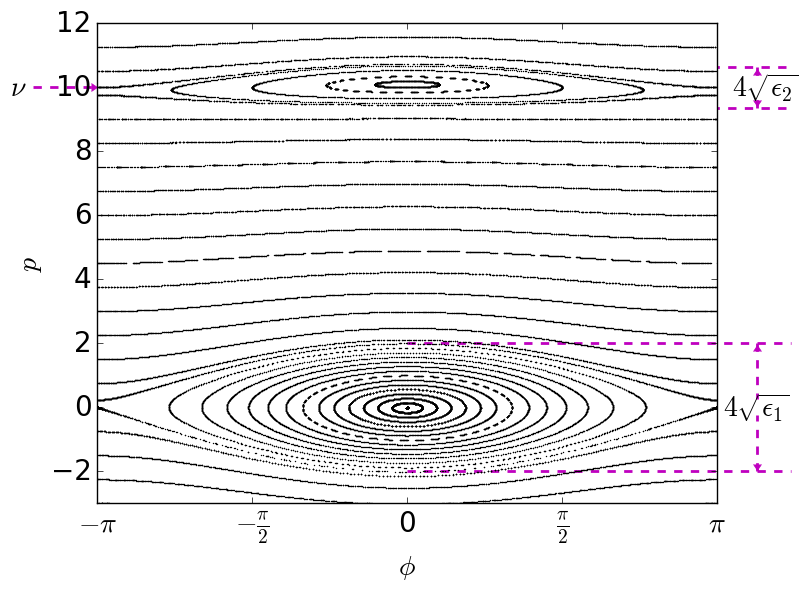}
    \centering
    \caption{A surface of section for the perturbed pendulum with $r = 1$, $\epsilon_1 = 1$, $\epsilon_2 = 0.1$, and $\nu = 10$.
            At this large value of $\nu$ the two resonances appear  as ``cat's eyes,'' and are clearly separated.
            There is also little chaos.}
    \label{sos_pict}
\end{figure}

% Surface of Section \[{
We start by using a surface of section to exhibit the dynamics of Eq. (\ref{toy_model}), similar to what is done in \cite{2001_murray} and \cite{2007_zaslavsky}.
For each surface of section, we fix the parameters ($r$, $\nu$, $\epsilon_1$ and $\epsilon_2$),
    and numerically integrate the perturbed pendulum's equations of motion for different initial values of $p$ and $\phi$.
This experiment is equivalent to fixing the orbits and masses of the two planets and exploring what happens for different initial semimajor axes of the test particle.
From each integrated trajectory, we plot snapshots in the $p$-$\phi$ plane at times $t$ such that $r\nu t = \set{0, 2\pi, 4\pi, \cdots}$.
Stable trajectories will create 1D curves in the $p$-$\phi$ plane, while chaotic ones will randomly fill a 2D region.
Figure \ref{sos_pict} shows an example surface of section with $r=1$,
    $\epsilon_1=1$, $\epsilon_2=0.1$, and a large forcing frequency ($\nu\gg 1$). 
In that case, each cosine term (i.e., ``resonance'') in Eq. (\ref{toy_model}) produces a ``cat's eye'' in the surface of section.
The $\epsilon_1$-resonance is centered at $p = 0$, and the $\epsilon_2$-resonance is centered at $\nu$.
Within the $\epsilon_1$-resonance, $\phi$ librates, and within the other resonance, it is $\psi$ that librates.
% }\]

\begin{figure}[!t]
    \includegraphics[width=0.99\columnwidth]{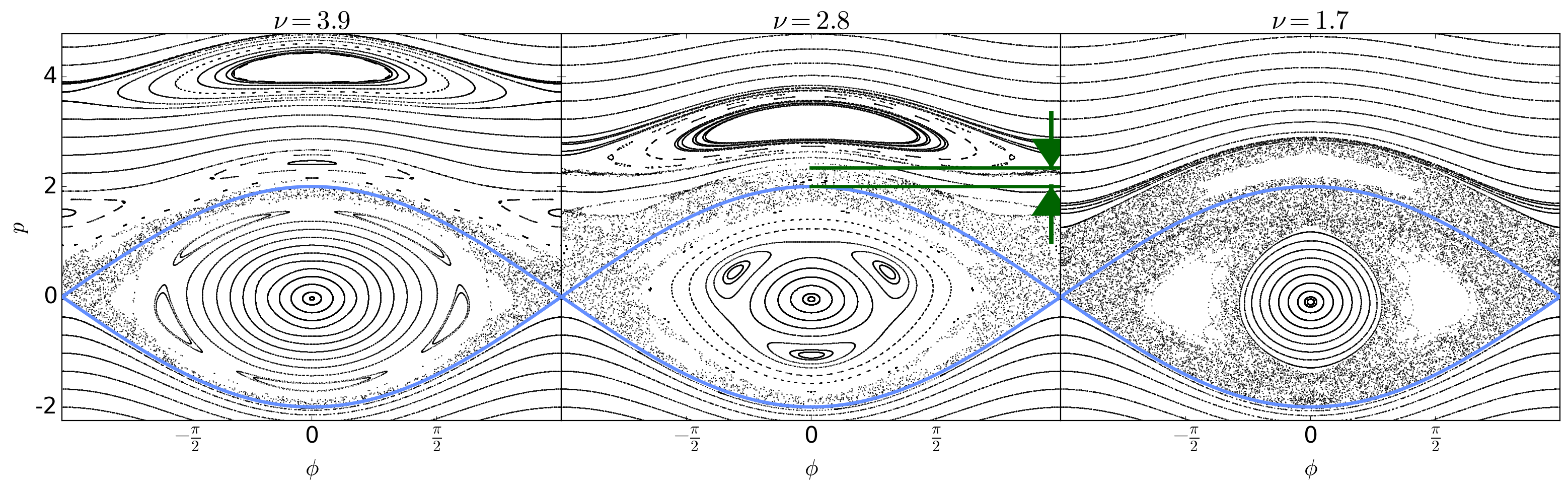}
    \centering
    \caption{Surfaces of section for the perturbed pendulum, as in Fig. \ref{sos_pict}, but for three lower values of $\nu$.
    t         The separatrices of the $\epsilon_1$ resonance are shown in blue.
             In the middle panel, the extent of the chaotic region predicted by Eq. (\ref{eq:ckc}) is indicated in green.}
    \label{sos}
\end{figure}

% Middle/Right Panel \[{
For Fig. \ref{sos}, we dial down $\nu$, which lowers the $\epsilon_2$-resonance in the $p$-$\phi$ plane.
In the context of the orbital problem, this corresponds to moving the planets so that the two resonances approach each other.
At moderate separation ($\nu = 3.9$), one observes new islands close to the lower resonance's separatrix, and the separatrix becomes chaotic.
As $\nu$ gets smaller, additional islands appear, and the chaotic region expands.
The $\epsilon_2$-separatrix also becomes chaotic.
But chaos does not extend below the lower $\epsilon_1$ separatrix, nor above the upper $\epsilon_2$ separatrix.
We quantify all of this behavior in the following sections.
% }\]

% Figure: Perturbed Pendulum \[{
% }\]

\begin{figure}[!t]
    \includegraphics[width=0.99\columnwidth]{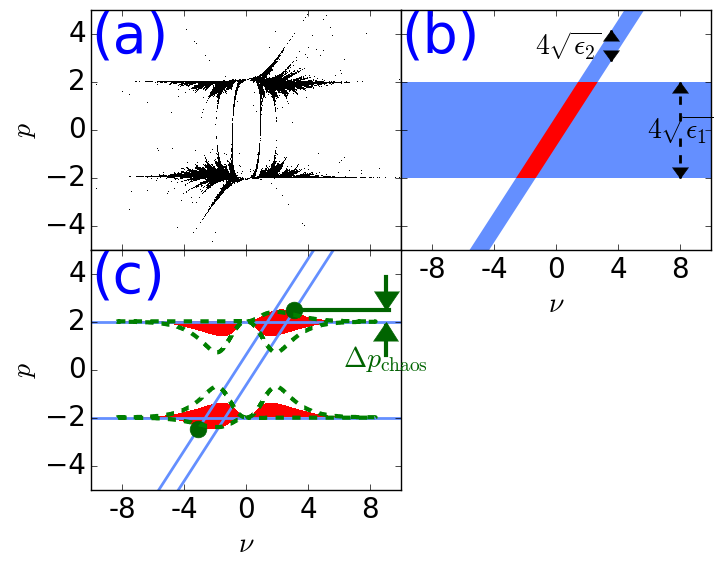}
    \centering
    \caption{(a) Double surface of section for the system in Figure \ref{sos},
                 showing where the chaotic region is for many different values of $\nu$ (at $\phi = 0$).
             (b) Prediction for (a) based on naive resonance overlap.
                 The libration regions of both resonances are blue, and the predicted chaotic region is in red.
             (c) The region predicted to be chaotic by the classical kick criterion (Eq. \ref{eq:ckc0}) is outlined with green dashed curves.
                 The improved kick criterion (Eqs. \ref{Ecrit0} \& \ref{jeremysDiscovery}) is shown with a red region.
                 The green circles show the reduced classical criterion (Eq. \ref{eq:peak}).}
    \label{ni}
\end{figure}

% Double Surface of Section \[{
We combine the plots for all values of $\nu$ by taking a ``double surface of section'':
    at each $\nu$ we take a cut in the surface of section at $\phi = 0$,
    and record the values of $p$ where trajectories are chaotic.
We then stack these for different values of $\nu$.
The result is shown in Fig. \ref{ni}(a).\footnote{
In practice, we make a double surface of section 
as follows: we initialize $\phi=\psi=0$,
    and $p=p_i$, and plot black in the $p_i$-$\nu$ plane wherever
    an orbit with those initial values are chaotic, as indicated by MEGNO.
    Note that the perturbed pendulum has 1.5 degrees of freedom, 
    and hence the boundary between chaotic and non-chaotic regions is well defined \cite[\S 4.6 in][]{1979_chirikov}.\label{foot:dsos}}
A double surface of section is analogous to the PP-plane, because a simple linear change of co-ordinates brings $\set{p, \nu}$ to $\set{P_1, P_2}$ (\S \ref{sec:we}).
Fig. \ref{ni}(b) shows, in blue, the locations of the two resonances in the $p$-$\nu$ plane.
This plot is the analog of Fig. \ref{kepler-11ro}.
The most naive criterion for chaos---resonance overlap---predicts that chaos should occur in the red zone of this plot.
Comparing with panel (a), we see that, as before, this prediction provides only a rough guide to the numerical result.
It does not explain, e.g., why most of the chaos occurs when the two resonances marginally overlap (i.e. at $\set{\nu,p}\sim \set{2, 2}$ and $\sim \set{-2, -2}$),
    nor why much of the chaos consists of spikes that extend from near the points of marginal overlap.
% }\]

% Figure: Surface of Sections \[{
% }\]

% Developing Theories \[{
In this paper, we develop two theories to explain the chaos in double surfaces of section,
    and hence in the PP-plane.
The first, which we call the theory of kicks, predicts the smoothed shape for the chaotic zones,
    e.g., as shown in Fig. \ref{ni}(c) in red.
The second is based on secondary resonance overlap.
It predicts the same smoothed shape for the chaotic zones.
But it also accounts for the spikes, and predicts the Lyapunov times to moderate accuracy.
% }\]

% Figure: Numerical Integration \[{
% }\]

\section{Theory of Kicks}\label{sec:theory}
% Describe Plan for Section \[{
In this section, we first derive the criterion for chaos based on the theory of kicks (\S \ref{sec:km}),
    which is similar to the ``whisker map'' of \cite{1979_chirikov} and the ``separatrix map'' of \cite{2007_zaslavsky}.
We then complete the derivation, leading to two explicit,
    but approximate, forms for the criterion (\S \ref{sec:akc}).
We call the first form the classical theory \citep{1979_chirikov, 1983_lichtenberg, 2007_zaslavsky},
    and the second incorporates our improvements.
Next, we apply the improved theory to a single resonance crossing from Fig. \ref{kepler-11},
    and finally apply it to many more resonance crossings in Fig. \ref{kepler-11}.
% }\]

\subsection{Kick Criterion}\label{sec:km}
% Introduce Kick Method \[{
We model the perturbed pendulum (Eq. \ref{toy_model}) as a simple pendulum subjected to discrete kicks that are given by the $\epsilon_2$ term in the Hamiltonian.
We focus on an unperturbed trajectory near the  separatrix (i.e., the $\epsilon_1$-separatrix).
For definiteness, we assume at first that $\nu > 0$ and the trajectory is above the separatrix, in the circulation zone. 
Whenever $\phi$ swings from  $-\pi$ to $\pi$ (or more generally from $(2k-1)\pi$ to $(2k+1)\pi$ for integer $k$), the pendulum is considered to have experienced a kick.
% }\]

% Get Criterion in Words \[{
Let us consider two successive kicks.
The key quantity of interest for whether or not chaos occurs is the {\it kick phase} $\psi = r(\phi-\nu t)$,
    which is the argument of the $\epsilon_2$-cosine.
Chaos is postulated to occur if the first kick is sufficiently strong that it changes the kick phase at time of the second kick---relative to what it would have been without the first kick---by $\gtrsim 1$.
If that happens, it means that the first kick was strong enough to scramble the amplitude of the second kick  ($\propto \cos\psi$)
by an order-unity amount. The scrambled second kick will in turn scramble 
the third kick, etc., leading to pseudo-random evolution and hence chaos.
% }\]

% Introduce Change in Energy \[{
To calculate that change in kick phase, we first determine how the first kick changes the unperturbed energy, $E = {p^2\over 2} -\epsilon_1\cos\phi$.
The equations of motion from the perturbed pendulum give
    \begin{align}
    {\td E\over \td t} = -\epsilon_2 r \dot{\phi}\sin(r(\phi - \nu t)) \ . \label{Edot}
    \end{align}
Since we consider $\epsilon_2$ to be small, we evaluate the change in energy from the first kick by integrating Eq. (\ref{Edot}) from half a period before the kick to half a period after, while taking $\phi(t)$ as unperturbed:\footnote{
    This expression for $\delta E$ is slightly incorrect because it includes ``fast oscillations,''
        whose effect average out.
    In Appendix \ref{app:ikc}, we show how the fast oscillations should be removed.}
    \begin{eqnarray}
    \delta E &=& -\epsilon_2 r \int_{t^{(1)}-{T\over 2}}^{t^{(1)}+{T\over 2}} \dot{\phi}_{\rm unp}\sin\left[ r\left(\phi_{\rm unp}+\phi(t^{(1)})-\nu t\right)\right]\td t \nonumber\label{eq:delta_e}
    \end{eqnarray}
    where $t^{(1)}$ is the time of the first kick;
    the unperturbed $\phi(t) = \phi_{\rm unp}(t - t^{(1)}) + \phi(t^{(1)})$, after adopting
    the convention that $\phi_{\rm unp}(t = 0) = 0$; and $T$ is the unperturbed period.
The above equation may be rewritten as
    \begin{align}
    \delta E &= -\epsilon_2 r \sin\psi^{(1)}{\cal K}(E) \label{eq:dek}
    \end{align}
    where
    \begin{equation}
    {\cal K}(E) = \int_{-{T\over 2}}^{{T\over 2}} \dot{\phi}_{\rm unp}(t')\cos\left[ r\left(\phi_{\rm unp}(t') - \nu t'\right)\right]\td t' \ , \label{calK}
    \end{equation}
    after changing variables to $t' = t - t^{(1)}$,
    making use of the symmetry of $\phi_{\rm unp}$ under time reversal,
    and defining $\psi^{(1)} = \psi(t = t^{(1)})$.
Approximate expressions for ${\cal K}(E)$ are presented below.
In addition to its dependence on $E$, ${\cal K}$ depends on the parameters $\epsilon_1$, $r$, and $\nu$.
% }\]

% Derive Criterion \[{
Given the change in energy, the period changes by  $\delta T={dT\over dE}\delta E$,
    where the $T(E)$ relationship for an unperturbed pendulum is well-known \citep[e.g.,][]{1983_lichtenberg}.
Therefore, the change in kick phase at the second kick is
    \begin{align}
    \delta \psi^{(2)} = -r\nu \delta T = \epsilon_2 r^2\nu\sin\psi^{(1)}~{\cal K}(E)\frac{\td T}{\td E}\ , \label{dpsi2}
    \end{align}
which has a ``typical'' value of 
    \begin{align} \delta \psi^{(2)}\sim \epsilon_2 r^2\nu{\cal K}(E)\frac{\td T}{\td E} \end{align}
Finally, for chaos to occur one requires this typical value to be $\gtrsim 1$;
    i.e., trajectories with energy $E$ that satisfy
    \begin{align}
    \left|\frac{\td E}{\td T}\right| \lesssim \epsilon_2 r^2\left|\nu{\cal K}(E)\,\right| \label{Ecrit0}
    \end{align}
    are chaotic.
We call this the kick criterion.
% }\]

\subsection{Two Approximate Kick Criteria}\label{sec:akc}
\subsubsection{Classical Kick Criterion}\label{sec:ckc}
% Discuss Separatrix K \[{
Previous studies \citep[e.g.,][]{1979_chirikov, 1983_lichtenberg, 2007_zaslavsky} take the unperturbed trajectory to be very near the separatrix $(E\sim \epsilon_1)$.
One may then evaluate ${\cal K}$ by setting $\phi_\t{unp}$ in Eq. (\ref{calK}) to its value on the separatrix, 
    and correspondingly set the integration limits to $\pm\infty$.
As we show in Appendix \ref{sec:MA}, the result is
    \begin{align}
    {\cal K} \approx \nu A_{2r}(r\nu) \label{K_chi}
    \end{align}
    where $A_{2r}(r\nu)$ is the Melnikov-Arnold (MA) integral,
    the value of which is given in Eq. (\ref{MA2}).
The resulting ${\cal K}$ is independent of $E$.
In deriving Eq. (\ref{K_chi}), we assume, as above, that the unperturbed trajectory is in the circulation zone;
    and we also set $\epsilon_1 = 1$ for clarity.
We shall continue to set $\epsilon_1 = 1$ in the remainder of this subsection,
    before generalizing to arbitrary $\epsilon_1$ at the end.
% }\]

% Derive Classical Criterion \[{
Equation (\ref{Ecrit0}) may be further simplified by using the near-separatrix relation $|dE/dT|\approx |E-1|$,
    resulting in the ``classical kick criterion:''
    \begin{align}
    |E - 1|< \epsilon_2 r^2 \nu^2 \left|A_{2r}(r\nu)\,\right| \ . \label{eq:ckc0}
    \end{align}
In order to depict this on a double surface of section, we
convert $E$ to the value of $p$ at $\phi = 0$,
    via $E = {p^2 \over 2} - 1 \approx 1 + 2\Delta p$,
    where $\Delta p$ is the height above the separatrix at $\phi = 0$.
In that case, Eq. (\ref{eq:ckc0}) becomes
    \begin{eqnarray}
        \Delta p < \Delta p_{\rm chaos,\nu} \equiv {1\over2}\epsilon_2 r^2\nu^2\left|A_{2r}(r\nu)\,\right|\ . \label{eq:ckc}
    \end{eqnarray}
The extent of the chaotic zone above the separatrix, $\Delta p_{\rm chaos,\nu}$,
    is depicted by arrows in Fig. \ref{sos}(b).
That extent depends on the height of the $\epsilon_2$ resonance, i.e., on $\nu$.
In particular, in Fig. \ref{ni}(c), the top green dashed curve in the upper-right quadrant shows the prediction of Eq. (\ref{eq:ckc}) for how $\Delta p_{\rm chaos,\nu}$ depends on $\nu$.
Comparing that part of the plot with the numerical result in panel (a) shows good agreement.
For $\nu\gg 1$, $A_{2r}(r\nu) {\propto} e^{-\pi r\nu/2}$ (Eq. \ref{MA2}),
    and so $\Delta p_{\rm chaos,\nu}$ decays exponentially as $\nu$ increases, as seen in panel (c).
Furthermore, $\Delta p_{\rm chaos,\nu}$ hits a peak value at $\nu\sim 1$,
    and then decays to zero as $\nu\rightarrow 0$.
The peak value is
    \begin{align}
    \Delta p_{\rm chaos} = {1\over 2} \epsilon_2 \kappa(r) \label{eq:peak}
    \end{align}
    where
    \begin{align}
    \kappa(r)\equiv \max_\nu r^2\nu^2\left|A_{2r}(r\nu)\,\right|
    \approx \begin{cases} 3.5r^{1.0} & r < 0.28\\ 8.9r^{1.8} & r > 0.28 \end{cases}\ \ ,
    \end{align}
    and where the latter approximation is from a simple fit that is in error by $\sim 20\%$ near the crossover (at $r = 0.28$), and by $\lesssim 5\%$ away from the crossover.
For the case shown in Fig. \ref{ni}, $r=1$ and so $\Delta p_{\rm chaos}={1\over 2} \times 0.1 \times 8.9= 0.44$.
In Fig. \ref{ni}(c) we plot a green circle at this distance from the separatrix; note that we position the $\nu$ of the circle along the $\epsilon_2$-resonance rather than at the peak of $\Delta p_{\rm chaos}$, 
    for reasons that will be described in \S \ref{sec:ikc}.
We refer to Eq. (\ref{eq:peak}) as the ``reduced classical criterion.''
It will prove useful for estimating the extent of chaotic zones in a PP diagram.
% }\]

% Describe Other Boundaries \[{
The remaining green dashed curves in Fig. \ref{ni}(c) are calculated similarly.
At $\nu>0$, there are three additional boundaries:
    in the upper libration zone (i.e., just below $p=2$),
    the lower libration zone, and the lower circulation zone.
For the first of these, the analysis is the same as previously, except that the period is twice as long,
    which implies that $|\Delta p_{\rm chaos,\nu}|$ should be increased by a factor of two there.
That factor of two is erroneously omitted in the literature \citep[e.g.,][]{1979_chirikov}.
The boundary in the lower libration zone must be the same as that in the upper,
    because the two regions depict the same orbits.
And in the lower circulation zone, one should simply switch $\nu\rightarrow -\nu$ in Eq. (\ref{eq:ckc})---because an orbit in the lower circulation zone with $\nu > 0$ is equivalent to one in the upper circulation zone with $\nu < 0$, and the previous analysis applies to the upper circulation zone.  
The remainder of the green curves, i.e., at $\nu < 0$,
    may be trivially obtained from those at $\nu > 0$ by  antisymmetry.
As seen in Fig. \ref{ni}, the predicted green curves all agree  moderately well with the numerical result of panel (a).
% }\]

% Generalize for e_1 /= 1 \[{
We conclude this subsection by generalizing Eqs. (\ref{K_chi}), (\ref{eq:ckc0}), and (\ref{eq:peak}) to arbitrary $\epsilon_1\ne 1$,
    which may easily be done by applying the rules in footnote \ref{ftn:cov} (as well as $E \rightarrow E/\epsilon_1$).
The results are as follows
    \begin{align}
    {\cal K} &\approx {\nu\over\sqrt{\epsilon_1}}A_{2r}\p{{r\nu\over\sqrt{\epsilon_1}}}
    \label{eq:kkk}
    \\
    \left|E - \epsilon_1\right| &< {\epsilon_2\over\epsilon_1}r^2\nu^2\left|A_{2r}\p{\frac{r\nu}{\sqrt{\epsilon_1}}}\right|
    \label{eq:ckc1} \ \ \ \rm classical
    \\
    \Delta p_{\rm chaos} &= {1\over 2}\frac{\epsilon_2}{\sqrt{\epsilon_1}}\kappa(r) \label{eq:peak2}
    \ \ \ \ \ \ \ \rm reduced\ classical
    \end{align}
    and $\kappa(r)$ is left unchanged.

% }\]

\subsubsection{Improved Kick Criterion:}\label{sec:ikc}
% Discuss Improvements \[{
Although the classical criterion (Eq. \ref{eq:ckc1}) is adequate for predicting the chaos in the system shown in Fig. \ref{ni}, it is inadequate for many resonant crossings in the PP diagram.
There are two main shortcomings.
First, when the chaotic region extends beyond the $\epsilon_1$ separatrix by
    a distance larger than the width of that resonance, 
    one may not set $\phi_{\rm unp}$ to its value on the separatrix.
From Eq. (\ref{eq:peak2}), that occurs when $\epsilon_2\kappa(r)\gtrsim \epsilon_1$.
And second, one must consider the chaos of the $\epsilon_2$-separatrix,
    which becomes dominant when $\epsilon_2\gtrsim \epsilon_1$.
In the following, we address these two shortcomings in turn.
% }\]

% Improved Kick Criterion \[{
For the first, we generalize the classical criterion by
    (a) evaluating $dE/dT$ in Eq. (\ref{Ecrit0}) with the full expression that is valid for any $E$;
    and (b) approximating ${\cal K}$ without assuming that the trajectory is on the separatrix.
We derive that improved approximation in Appendix \ref{app:ikc}.
The derivation is subtle, because one must distinguish between fast and slow oscillations.
But the final result is simple: in Eqs. (\ref{K_chi}) and (\ref{eq:kkk}),
    one should replace $\nu\rightarrow \nu-\Delta p$,
    where $\Delta p$ is the height above the separatrix at $\phi = 0$,
    i.e., $\Delta p\equiv \sqrt{2(E+\epsilon_1)} - 2\sqrt{\epsilon_1}$.
Physically, that result may be understood from the principle that the strength of the kick should depend on the difference in frequency between the perturbing $\epsilon_2$ resonance and that of the test particle at the time of the kick (see Appendix \ref{app:ikc}).
It is for this reason that we position the green circles in Fig. \ref{ni} (Eq. \ref{eq:peak}) along the $\epsilon_2$ separatrix:
    the maximum extent of chaos should occur when $\nu - \Delta p \sim 1$ rather than $\nu \sim 1$.
% }\]

% K Improved \[{
The resulting improved approximation for ${\cal K}$ is given by Eq. (\ref{K_mine}), for the case when $\nu>0$,
    $\epsilon_1=1$ and the trajectory is in the upper circulation zone.
That is easily generalized to arbitrary $\epsilon_1$ and to the remaining zones,
    following what we did for the classical criterion.
To be explicit, the final ``improved kick criterion'' is given by Eq. (\ref{Ecrit0}), in which
    \begin{align}
    {\cal K}(E) \approx \begin{cases} \frac{(\nu - \left|\Delta p\right|)}{\sqrt{\epsilon_1}} A_{2r}\left[\frac{r(\nu - \left|\Delta p\right|)}{{\sqrt{\epsilon_1}}}\right] & \t{upper circulation}\\ \frac{(\nu + \left|\Delta p\right|)}{{\sqrt{\epsilon_1}}} A_{2r}\left|\frac{r(\nu + \left|\Delta p\right|)}{{\sqrt{\epsilon_1}}}\right| & \t{libration}\\ \frac{(\nu - \left|\Delta p\right|)}{{\sqrt{\epsilon_1}}} A_{2r}\left[-\frac{r(\nu - \left|\Delta p\right|)}{{\sqrt{\epsilon_1}}}\right] & \t{lower circulation} \end{cases} \label{jeremysDiscovery}
    \end{align}
   when $\nu > 0$.
Now ${\cal K}$ depends on $E$ via its dependence on $\Delta p$. 
Very near the separatrix ($E\sim \epsilon_1$), ${\cal K}$ reduces to its classical expression.
The case when $\nu < 0$ is found by antisymmetry.
We show the improved criterion in red in Fig. \ref{ni}(c).
Its agreement with the numerical result in panel (a) is slightly better than the classical prediction,
    particularly inside the libration zone.
% }\]

% Figure: Analytic Theory Illustration \[{
\begin{figure}[!t]
    \includegraphics[width=0.995\columnwidth]{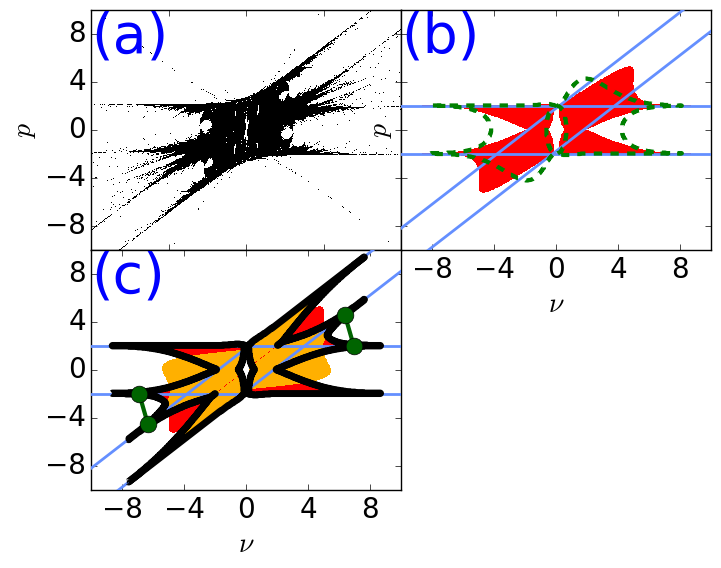}
    \centering
    \caption{A more chaotic system:
             (a) Numerical double surface of section,
                 for a system the same as in Fig 5, but $\epsilon_2$ increased to 0.8.
             (b) The predicted chaotic region for the $\epsilon_1$ resonance using the classical criterion
                 (green outline) and the improved criterion (red region).
             (c) Chaos from both resonances: the improved criterion for the $\epsilon_1$ resonance is in red,
                 copied from panel (b), and that for the $\epsilon_2$ resonance is overplotted in orange.
                 The merged prediction is outlined in black.
                 The green circles show the reduced classical criterion,
                    now applied to both the $\epsilon_1$ and $\epsilon_2$ resonances.}
    \label{reverse_prediction}
\end{figure}
% }\]

% Discuss Figure 6 Initially \[{
In Fig. \ref{reverse_prediction}(a)--(b), we repeat the study from Fig. \ref{ni},
    but now with $\epsilon_2$ increased from 0.1 to 0.8, in which case $\Delta p_{\rm chaos}=3.6$.
That is comparable to the width of the $\epsilon_1$ resonance ($\Delta p=4$),
    and so one expects the classical criterion to be inadequate.
The numerical chaotic region is shown in panel (a), and it is now seen to be much larger than before.
In fact, it  agrees much better with the prediction of naive resonance overlap, i.e.,
    with the analog of  Fig. \ref{ni}(b) after increasing to $\epsilon_2=0.8$.
That fact is generally true: naive resonance overlap works best when the widths of the two overlapping resonances are comparable \citep[e.g.,][]{Shepelyansky_2009}.
% }\]

% Address Numerical Chaotic Region \[{
In panel (b), we show the predictions from the classical criterion (green dashed curve) and from the improved criterion (red region).
The red region is a much better match to the numerical result.
In particular, the red region predicts that chaos in the circulation zone roughly follows the $\epsilon_2$ separatrix, whereas the classical prediction does not.
% }\]

% Include Epsilon_2 Resonance \[{
We turn now to the chaos of the $\epsilon_2$ separatrix.
To plot its improved kick criterion we may reuse the prior results,
    but with the roles of the two resonances swapped.
We proceed by first writing the Hamiltonian as
    $H(\phi_{sw}, p_{sw}, t) = {p_{sw}^2/2 - \epsilon _ 2\cos(\phi_{sw}) - \epsilon_1\cos(r_{sw}(\phi_{sw} - \nu_{sw} t))}$,
    where the ``sw'' subscript means those quantities are defined with 1 and 2 indices swapped in Eqs. (\ref{eq:rdef})--(\ref{eq:deflast}).
We then use the chaos criterion to determine the chaos threshold in the $p_{sw}$-$\nu_{sw}$ plane.
And finally, we map that into the $p$-$\nu$ plane via the transformation $p = p_{sw} + \nu_{sw}$ and $\nu = -\nu_{sw}$.
For the system in Fig. \ref{reverse_prediction}, the result is shown panel (c) in orange.
% }\]

% Merge Prediction \[{
Finally, we must merge the red and orange regions, which we do as follows:
    we discard the red when it lies inside of the $\epsilon_2$ resonance,
    because there the $\epsilon_2$ resonance dominates the dynamics;
    and similarly, we discard the orange when it is within the $\epsilon_1$ resonance.
That procedure discards everything in the region where both resonances overlap
    (e.g. the red region in Fig. \ref{ni}(b)), which is incorrect: 
    in that region, the stronger resonance dominates the dynamics.
Therefore if $\epsilon_1 > \epsilon_2$ we keep the red in the overlapping region,
    and otherwise we keep the orange there.
The resulting prediction is outlined in black in panel (c).
It is seen to agree quite well with the numerical result.
We also show in panel (c) the green circles from the reduced classical criterion (Eq. \ref{eq:peak}),
    but now for both of the resonances.
% }\]

\subsection{A Worked Example: The 5:3--2:3 Crossing}\label{sec:we}
% Figure: Surface of Section for new Case \[{
\begin{figure}[!t]
    \includegraphics[width=0.95\columnwidth]{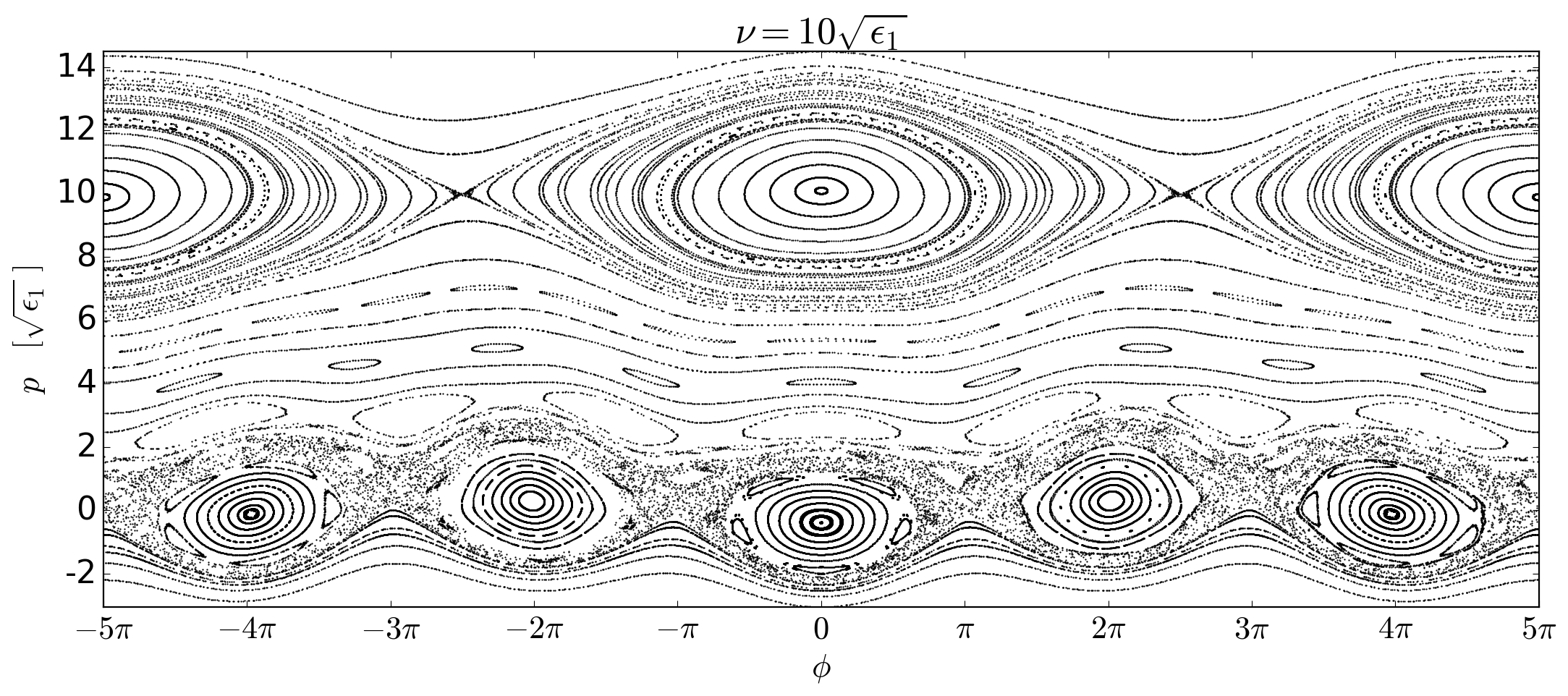}
    \centering
    \caption{Surface of section of the perturbed pendulum,
                with the parameters of the 5:3--2:3 resonance crossing, and $\nu$ as shown.}
    \label{workedsos}
\end{figure}
% }\]

% Introduce Specific Case \[{
We use our prior results to predict the shape of the chaotic zone at the crossing between the 5:3 and 2:3 resonances in Fig. \ref{kepler-11}.
The resonant integers are $j_1=5$, $k_1=2$, $j_2=2$, and $k_2=1$, and hence $r=j_2/j_1=2/5$. 
The resonant strength from the inner planet is $\epsilon_1=\mu_1|C_1|\tilde{e}_1^2$ (Eqs. \ref{eq:epsdef} and Appendix \ref{sec:ecce}), where $\mu_1$ is the planet-to-star mass ratio, $\tilde{e}_1$ is roughly the relative eccentricity between the test particle and the inner planet (See Eq. \ref{ediff} for the precise definition), and $C_1$ is the Laplace coefficient for the 5:3.
For the parameters of Fig. \ref{kepler-11}, $\epsilon_1 \approx 9.5\times10^{-7}$.
A similar calculation yields $\epsilon_2 = \mu_2 C_2\tilde{e}_2 \approx 3.3\times10^{-6}$.
As explained in \S \ref{sec:rpp}, $\epsilon_1$ may be scaled out of the problem,
    in which case the strength of the perturbing resonance becomes $\epsilon_2/\epsilon_1$,
    i.e., it is only the relative strengths of the two resonances that determines which orbits are chaotic. 
For the present case, we have $\epsilon_2/\epsilon_1\approx 3.5$.
Although we retain dependencies on $\epsilon_1$ in this subsection for completeness,
    the following results may be equally well  understood by setting $\epsilon_1\rightarrow 1$ everywhere.
% }\]

% Discuss Figure 7 \[{
Fig. \ref{workedsos} shows a surface of section of Eq. (\ref{toy_model}),
    when the height of the $\epsilon_2$ resonance is $\nu = 10\sqrt{\epsilon_1}$.
It differs from the one shown in Fig. \ref{sos_pict} in two notable ways.
First, because $r=2/5$ the $\epsilon_1$ resonance now has five cat's eyes for every two of the $\epsilon_2$ resonance.
And second, the large value of $\epsilon_2/\epsilon_1$ produces chaos of the $\epsilon_1$ resonance even at this large separation.
% }\]

% Figure: Square of Toy Model \[{
\begin{figure}[!t]
    \includegraphics[width=0.95\columnwidth]{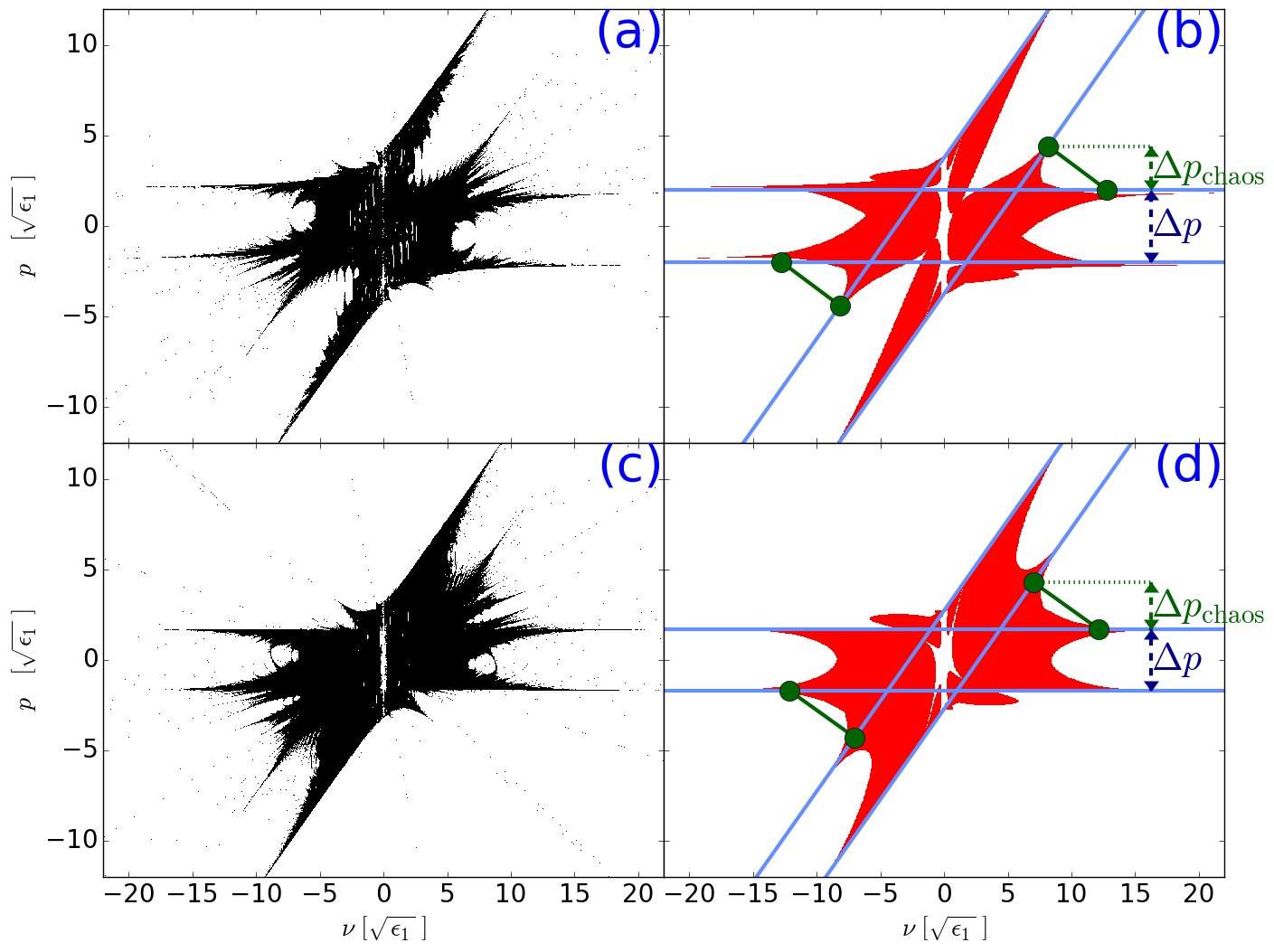}
    \centering
    \caption{Double surface of section of the perturbed pendulum,
                with the parameters of the 5:3--2:3 resonance crossing.
            (a) Numerical result. (b) Prediction of the improved kick criterion (red) and estimate from the reduced classical criterion (green circles). (c) and (d): Same as (a) and (b), but now accounting for the initial resonance angles.}
    \label{workedtoy}
\end{figure}
% }\]

% Discuss Figure 8 at Zero Phase \[{
Fig. \ref{workedtoy}(a) is the numerical double surface of section,
    and panel (b) is the prediction of the improved kick criterion.
    For the prediction, the chaotic regions of both resonances have been merged,
    as for the black outlined region in Fig. \ref{reverse_prediction}(c).
The prediction agrees well with the numerical result.
The green circles show the rough estimates from the reduced classical kick criterion---i.e.,
    from Eq. (\ref{eq:peak2}) for the top-most circle,
    and from the analogous equation for the right-most circle,
    which marks the chaos of the $\epsilon_2$ resonance.
Also shown is the width of the $\epsilon_1$ separatrix, $\Delta p = 4\sqrt{\epsilon_1}$.
% }\]

% Discuss Figure 8 at Non-Zero Phase \[{
The final step for comparing with Fig. \ref{kepler-11} is to transform Figs. \ref{workedtoy}(a)-(b) to the PP plane.
But before doing so, we must account for the initial resonance angles ($\phi_i$ and $\psi_i$),
    which are determined from the initial orbital elements via the expressions in footnote \ref{foot:phase} in Appendix \ref{sec:circ}.
We find $\phi_i \approx 1.1~\t{rad}$  and $\psi_i \approx 4.8~\t{rad}$.
By contrast, Fig. \ref{workedtoy}(a) is initialized at $\phi_i = \psi_i = 0$ (see footnote \ref{foot:dsos}).
Therefore panel (c) repeats (a), but now from integrations that start at the correct initial angles.
Correspondingly, the prediction in panel (d) is determined as in panel (b),
    but, whereas before we converted from $E\rightarrow p$ assuming $\phi=0$,
    we now use $p = \sqrt{2(E + \epsilon_1\cos \phi_)}$.
We use an analogous expression for the $\epsilon_2$ resonance.
% }\]

% Figure: PP Zoom in Plot \[{
\begin{figure}[!t]
    \centering
    \includegraphics[width=0.80\columnwidth]{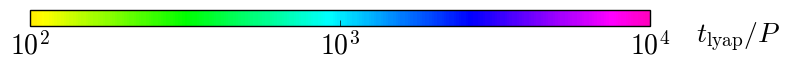}\\
    \includegraphics[width=1.00\columnwidth]{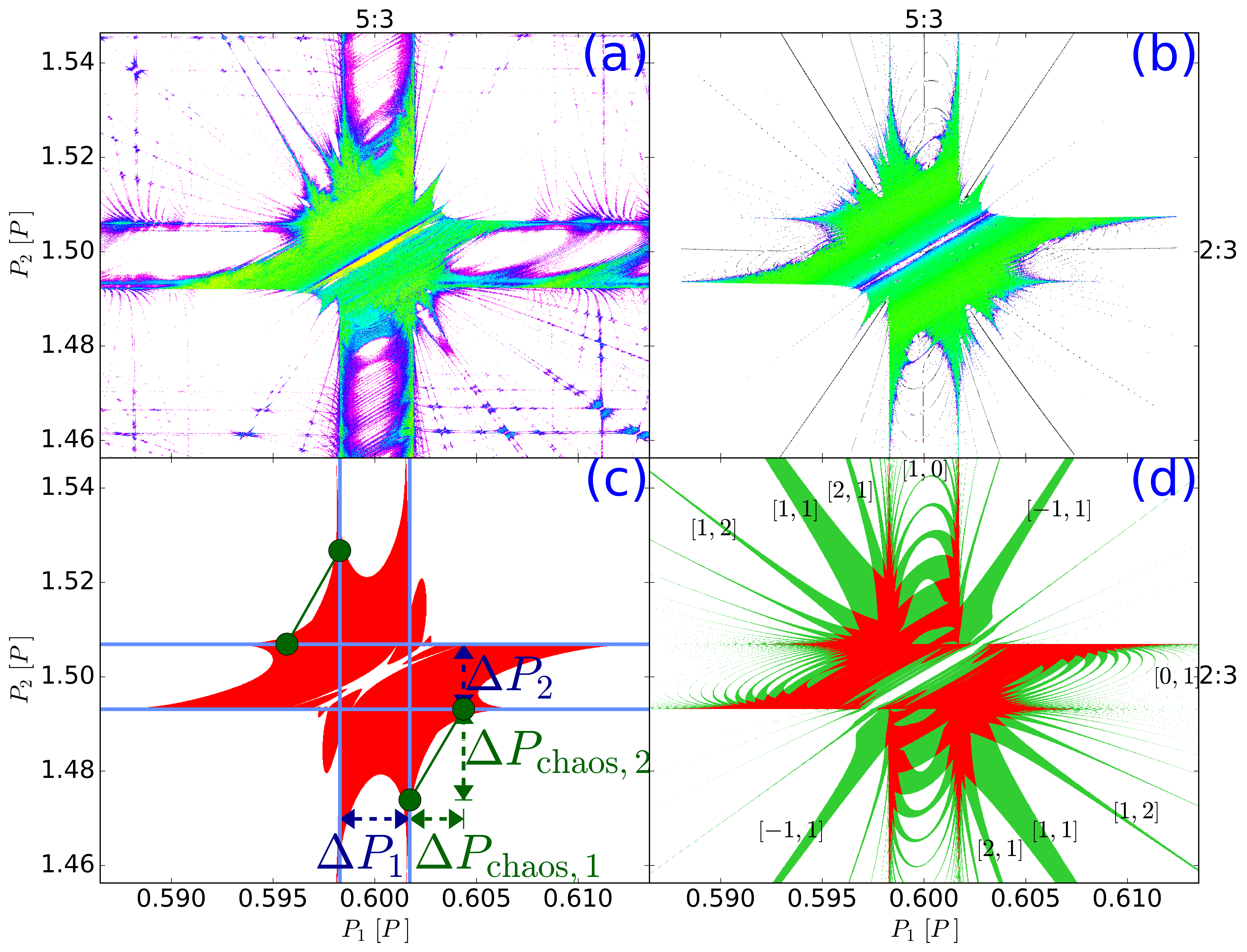}
    \caption{The 5:3--2:3 resonance crossing:
             (a) Lyapunov time in N-body simulations, in a zoomed-in region of Fig. \ref{kepler-11}.
             (b) Lyapunov time in integrations of perturbed pendulum model.
             This panel is the same as Fig. \ref{workedtoy}(c) after a simple coordinate transformation,
                and after a change in the color-coding to indicate Lyapunov time.
             (c) The analytic prediction of the improved kick criterion for the chaotic region.
             (d) The location of (inflated) secondary resonances (green) and where they overlap (red).}
    \label{workednbody}
\end{figure}
% }\]

% Equations for Figure 9 \[{
Fig. \ref{workednbody} shows the final result for this resonance crossing.
Panel (a) repeats the N-body simulations of Fig. \ref{kepler-11}, zoomed in to the 5:3--2:3 crossing.
Panel (b) repeats the numerical integration of the perturbed pendulum (Fig. \ref{workedtoy}(c)),
    with two alterations: first,
    the coordinates are changed from $p$-$\nu$ to those of the PP plane via the linear transformation
    \begin{align}
    \frac{P_1}{P} &\approx \frac{j_1 - k_1}{j_1}\pd{1 + \sqrt{3}p} \label{P1}\\
    \frac{P_2}{P} &\approx \frac{j_2 + k_2}{j_2}\pd{1 + \sqrt{3}\pd{p - \nu}}\ , \label{P2}
    \end{align}
    which follows from Eqs. (\ref{eq:nudef}) and (\ref{eq:deflast}) after expanding to linear order in $p$ and $\nu$.
And second, the chaotic region is colored to show the Lyapunov time.
Comparing panels (a) and (b) shows that the perturbed pendulum model explains the bulk of the N-body chaos,
    albeit not the very weak chaos with Lyapunov time $\gtrsim 10^3P$.
Panel (c) shows the prediction from the improved kick theory,
    which results from transforming the coordinates of Fig. \ref{workedtoy}(d).
The theory accounts for much of the structure seen in panels (a) and (b).
Also shown in panel (c) are the widths of the resonances,
    and estimates of the chaotic extents.
The width of the $\epsilon_1$ resonance follows from $\Delta p = 4\sqrt{\epsilon_1}$, Eq. (\ref{P1}),
    and accounting for the initial phase $\phi_i$ as described above, which together imply
    \begin{align}
    {\Delta P_1\over P} = 4\sqrt{3\epsilon_1}\ {j_1-k_1\over j_1}\p{{1+\cos\phi_i\over 2}}^{1/2}\ . \label{dPin}
    \end{align}
The extent of the $\epsilon_1$ separatrix  from the reduced classical criterion (Eq. \ref{eq:peak2}) is
    \begin{align}
    {\Delta P_{\rm chaos,1}\over P} \approx \frac{\sqrt{3}}{2}\ {j_1-k_1\over j_1}{\epsilon_2\over\sqrt{\epsilon_1}}\kappa\pd{r} \ . \label{PchaosI}
    \end{align}
For brevity, we ignore the dependence on $\phi_i$ in the latter expression,
    which is appropriate when $\Delta P_{\rm chaos,1}\lesssim \Delta P_1$.
For an average $\phi_i$, the ratio of the widths is $\Delta P_{\rm chaos,1}/\Delta P_1\sim 0.2{\epsilon_2\over\epsilon_1}\kappa(r)$, which evaluates to 1.2 for this system.
The extents of the $\epsilon_2$ resonance and chaotic region follow similarly, yielding
    \begin{align}
    {\Delta P_2\over P} &= 4\sqrt{3\epsilon_2}\ {j_2 + k_2\over j_2}\left({1+\cos\psi_i\over 2}  \right)^{1/2} \label{dPout}\\
    {\Delta P_{\rm chaos,2}\over P} &\approx \frac{\sqrt{3}}{2}\ {j_2 + k_2\over j_2}{\epsilon_1\over\sqrt{\epsilon_2}}\kappa\p{{1\over r}} \ . \label{PchaosO}
    \end{align}
The ratio of widths is $\Delta P_{\rm chaos,2}/\Delta P_2\sim 0.2{\epsilon_1\over\epsilon_2}\kappa\p{{1\over r}}\approx 2.3$.
% }\]

% Discuss Figure 9 (d) \[{
In sum, the improved kick criterion  explains the overall shape of the chaos seen in panels (a) and (b).
But it does not address the finer-scale structure.
As we shall show in \S \ref{sec:so} much of that is explained by the overlap of secondary resonances.
We show the final result in panel (d),
    but defer discussion of that panel to \S \ref{sec:apply}.
% }\]

% Figure: PP Plot w/ Theory \[{
\begin{figure}[!t]
    \includegraphics[width=1.0\columnwidth]{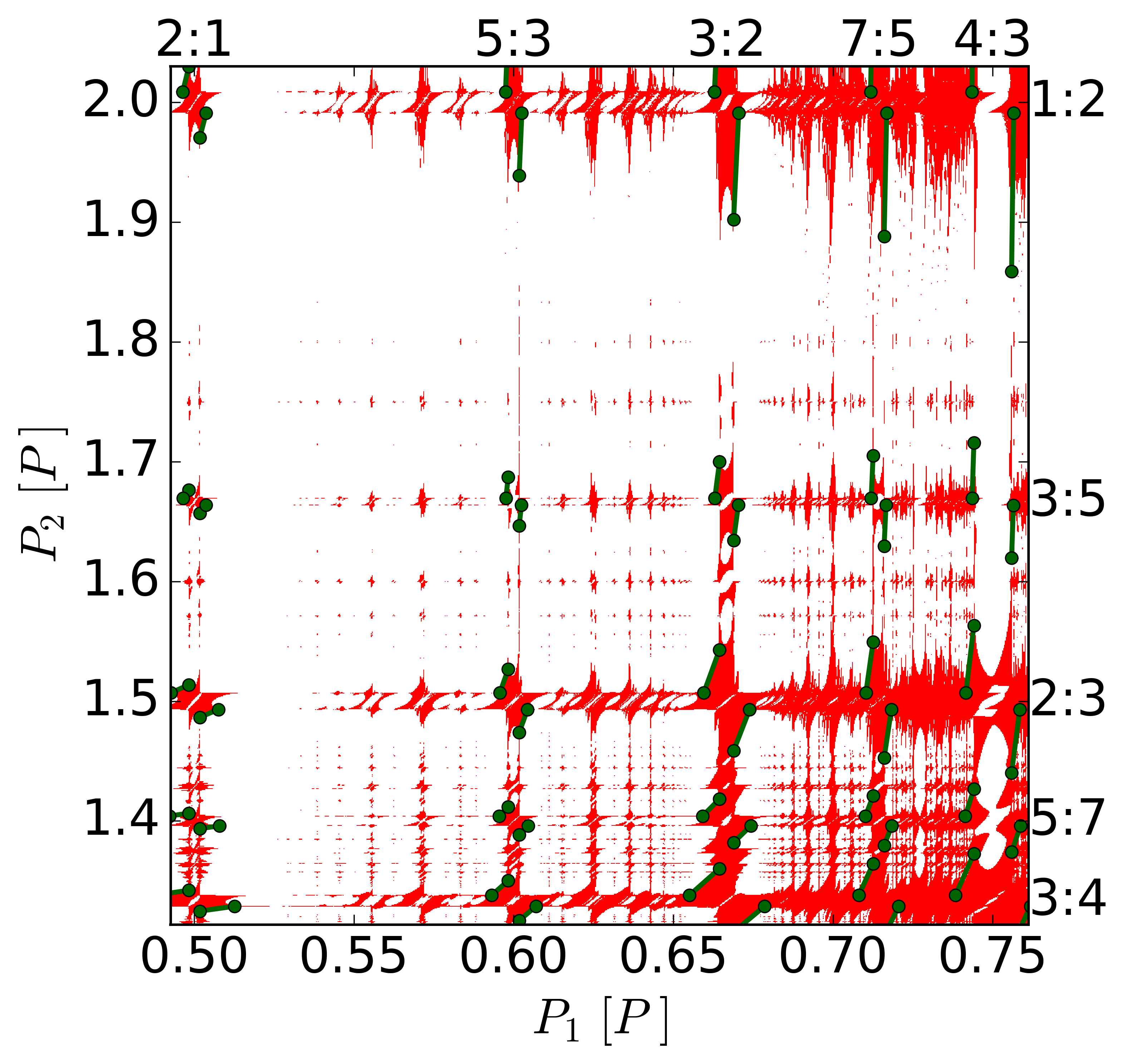}
    \caption{The red shaded region shows the prediction for where chaos should occur in Fig. \ref{kepler-11},
                 based on the improved kick criterion for resonances up to 10th order.
             The reduced classical prediction is shown as green circles for resonances up to 2nd order.}
    \label{original_cross_1}
\end{figure}
% }\]
~\\
\subsection{Many Crossings}\label{sec:fig1}
% Turn to PP Diagram \[{
We repeat the prior calculation for many more resonance crossings in Fig. \ref{kepler-11}.
Fig. \ref{original_cross_1} shows the prediction of the improved criterion in red,
    for resonances up to 10th order.
It reproduces the majority of the strong chaos seen in Fig. \ref{kepler-11},
    with Lyapunov time $\lesssim 10^4P$.
We also show the reduced classical criterion (green circles) for the largest resonance crossings.
These show that the classical estimate is an adequate approximation.
It improves upon naive resonance overlap when the two resonances have different sizes.
% }\]

\section{Theory of Overlapping Secondary Resonances}\label{sec:so}
% Motivate Section and Introduce Notation \[{
In surfaces of section such as Figs. \ref{sos} and \ref{workedsos},
    one observes that new cat's eyes (``secondary resonances'') appear when the two primary resonances are sufficiently close.
These secondary  resonances correspond to 3-body resonances in the orbital problem.
Here, we work out their {\it locations} and {\it widths}.
We also show that the overlap of neighboring secondary resonances reproduces the improved kick criterion. 
Finally, we use secondary resonances to estimate the Lyapunov times.
% }\]

\subsection{Locations and Widths of Secondary Resonances}\label{sec:wl}
% Expand Hamiltonian \[{
We analyze the perturbed pendulum Hamiltonian (Eq. \ref{toy_model}) as in standard Hamiltonian perturbation theory \cite[e.g.][]{1983_lichtenberg}.
First, the unperturbed part is rewritten in action-angle variables ($J, \theta$):
    \begin{align}
    H_{\rm pend}(\phi, p)\equiv {p^2\over 2} - \epsilon_1\cos\phi \ \ \ \rightarrow H_{\rm pend}(J) \ . \label{eq:unp}
    \end{align}
Hamilton's equations for the unperturbed motion now read $J = {\rm const}$, and $\dot{\theta} = {dH_{\rm pend}\over dJ}\equiv \omega = \t{const}$., where $\omega = 2\pi/{\rm period}$.\footnote{
    The inverse of the function $H_{\rm pend}(J)$ is given explicitly in Eq. (1.3.10) in \cite{1983_lichtenberg}.}
And second, the effect of the perturbation on an unperturbed trajectory with given $J$ is studied by expanding $H' = \epsilon_2\cos\left[r\pd{\phi - \nu t}\right]$ in a Fourier series in the angle, $\theta$.
Each term in the series will have a fixed frequency as the pendulum follows its unperturbed trajectory.
In Appendix \ref{sec:srd}, we show that the Fourier series is
    \begin{equation}
    H'(\theta,J,t) = \epsilon_2\sum_{M = -\infty}^\infty c_M(J)\cos\left[M\theta + r\pd{l\theta - \nu t}\right]  \ , \label{eq:hpert}
    \end{equation}
    where $l = 0$ when the unperturbed trajectory is in the libration zone,
    and $l = 1$ when it is in the circulation zone, and the Fourier coefficients are
    \begin{align}
    c_M(J) &= \frac{1}{2\pi}\int_{-\pi}^{\pi}\cos\pd{r\phi_{\rm unp}(\theta, J) - \pd{M + lr}\theta}\td\theta  \ .\label{cmexact}
    \end{align}
Note that we suppress functional dependencies on the  parameters ($\epsilon_1,\epsilon_2,\nu,r$).
In the integrand of Eq. (\ref{cmexact}), $\phi_{\rm unp}(\theta,J)$ describes the pendulum as it follows its unperturbed motion;
    it is the same function as appears in the kick criterion (Eq. \ref{calK}),
    except now we relabel its argument $t\rightarrow \theta=\omega t$,
    and we also make its dependence on $J$ explicit.
Equation (\ref{cmexact}) may be evaluated numerically,
    with $\phi_{\rm unp}(\theta, J)$ taken from \cite{1978_smith}.
An approximation for $c_M$ that is valid near the $\epsilon_1$ separatrix is derived in Eq. (\ref{cm4}) in Appendix \ref{sec:srd}, and one valid far from the separatrix is provided in Appendix D of \cite{1985_escande}.
% }\]

% Single M Term \[{
The perturbed pendulum Hamiltonian in action-angle variables is therefore $H(\theta, J, t) = H_{\rm pend}(J) + H'(\theta, J, t)$, without approximation.
Each cosine term in the Fourier series for $H'$ produces a secondary resonance.
If we focus on a single $M$ term, the Hamiltonian becomes
    \begin{align}
    H(\theta, J, t) \approx H_{\rm pend}(J) + \epsilon_2 c_M\cos\left[M\theta + r\pd{l\theta - \nu t}\right] \ . \label{secpend}
    \end{align}
The center of the resonance is where the argument of the cosine is stationary:
    $(M + rl)\dot{\theta} - r\nu = 0$.
After setting $\dot{\theta}=\omega$, as is valid to leading perturbative order,
    the $M$'th resonance occurs where the unperturbed pendulum has frequency
    \begin{align}
    \omega_M = {r\nu \over M + lr} \ . \label{eq:phys}
    \end{align}
% }\]

% Explain w_M \[{
To understand Eq. (\ref{eq:phys}), consider first the case that the unperturbed pendulum lies in the circulation zone ($l=1$).
Then the frequency of the $\epsilon_2$ term in Hamiltonian (\ref{toy_model}) is $r(\dot{\phi}-\nu)\sim r(\omega-\nu)$.
Equating that forcing frequency to an integer multiple of the unperturbed pendulum's frequency ($\omega$) reproduces Eq. (\ref{eq:phys}) (when the integer is $-M$).  
Likewise, if the unperturbed pendulum lies in the libration zone, then the forcing frequency is $r(\dot{\phi}-\nu)\sim -r\nu$, because $\dot{\phi}$ averages to zero.
Equating that to $-M\omega$ reproduces Eq. (\ref{eq:phys}) at $l=0$.
% }\]

% Discuss [M, N] Resonances \[{
General secondary resonances of the Hamiltonian in Eq. (\ref{toy_model}) have resonant angle in the circulation zone\footnote{
    Eq. (\ref{eq:mn}) is the resonant angle when the unperturbed $\phi$ and $\psi$ are circulating.
    When the unperturbed $\phi$ is librating and $\psi$ is circulating,
        the resonant angle is $(M-Nr)\phi+N\psi$.}
    \begin{equation}
    \Phi_{[M, N]} = M\phi + N\psi \label{eq:mn}
    \end{equation}
    for integers $M$ and $N$.
We call these $[M, N]$ resonances, in which case Eq. (\ref{eq:phys}) describes $[M, 1]$ resonances. 
In a slight abuse of terminology, we shall also label the $\epsilon_1$ resonance as the $[1,0]$ and the $\epsilon_2$ as the $[0,1]$.\footnote{
    It is an abuse because the argument of the $\epsilon_2$ cosine in the perturbed pendulum differs from that of the $M = 0$ cosine term in Eq. (\ref{eq:hpert}).
    The latter provides a slightly more accurate description of the $\epsilon_2$ resonance; or, to be more precise,
    Eq. (\ref{eq:hpert}) with $M = 0$ incorporates not only the $\epsilon_2$ resonance,
        but also its perturbation by the $\epsilon_1$.}
% }\]

% Figure: Secondary Overlap \[{
\begin{figure}[!t]
    \includegraphics[width = 0.99\columnwidth]{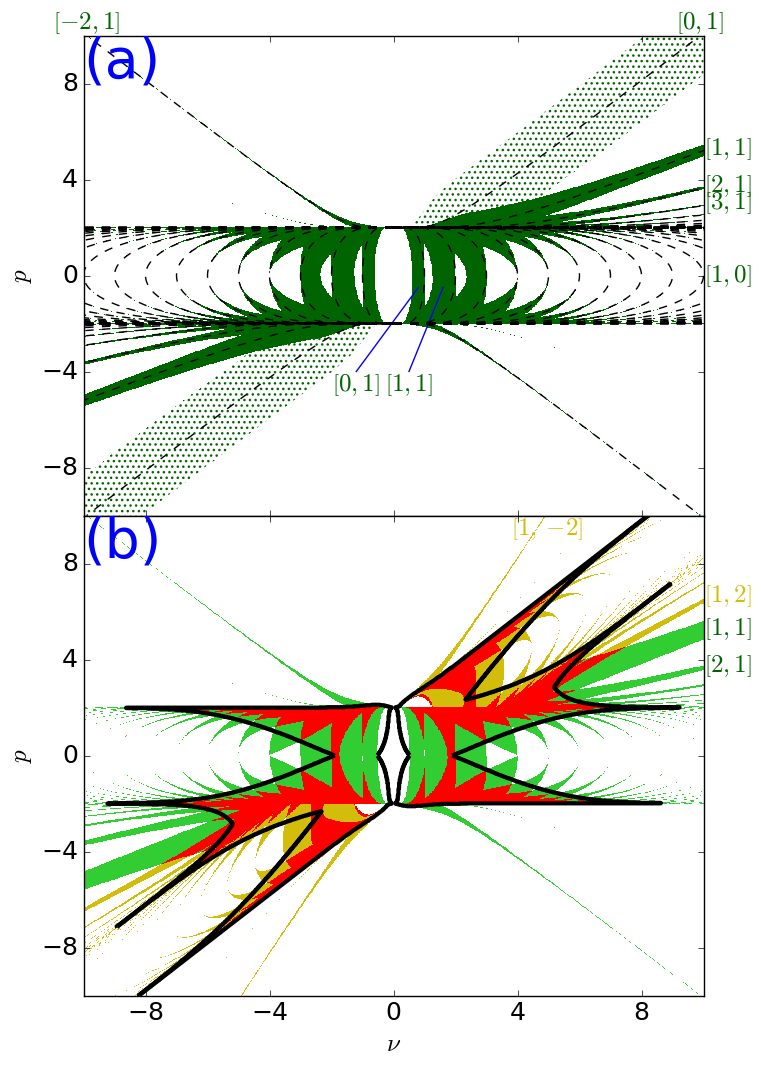}
    \centering
    \caption{Overlap of secondary resonances leading to chaos,
                for a perturbed pendulum with the same parameters as Fig. \ref{reverse_prediction}.
             (a) Locations of the $[M, 1]$ secondary resonances are shown as black dashed curves,
                and their widths are shaded green.
             The $[0, 1]$ is stippled because we neglect it in determining where overlap occurs.
             (b) $[M, 1]$ secondary resonances (light green) are copied from panel (a),
                after inflating by $\pi/2$ and excluding the $[0, 1]$.
             The yellow are the same for the $[1,N]$ resonances.
             The region where at least two secondary resonances overlap is shown in red. 
             The black curve is the prediction of the improved kick criterion,
                copied from Fig. \ref{reverse_prediction}(c).}
    \label{2nd_overlap}
\end{figure}
% }\]

% Describe Secondary Overlap Figure \[{
Equation (\ref{eq:phys}) determines the locations of the $[M,1]$ resonances.
To plot them in the $p$-$\nu$ plane,
    i.e., in the plane used for double surfaces of section,
    at each $\nu$ we use the properties of the unperturbed pendulum to convert from $\omega_M$ to energy $E$,
    and thence to the value of $p$ at $\phi=0$ (via $E = {p^2\over 2}-\epsilon_1$).
The result is shown in Fig. \ref{2nd_overlap}(a) as black dashed lines,
    for the same system as Fig. \ref{reverse_prediction}.
% }\]

% Secondary Widths \[{
We turn now to the widths of the resonances, which are determined by first Taylor expanding $H_{\rm pend}(J)$ in Eq. (\ref{secpend}) about $J_M$ (the action corresponding to $\omega_M$) which gives $H_{\rm pend}\approx {\rm const}+\omega_M(J-J_M)+\frac{1}{2}\left.\frac{\td^2 E}{\td J^2}\right|_{J = J_M}\pd{J - J_M}^2$.
The term linear in $J-J_M$ is removed by a canonical change of variables,
    and Eq. (\ref{secpend}) is turned into a simple pendulum by
    dividing through by $d^2E/dJ^2$ (and correspondingly rescaling time by the same factor).
The full-width can then be read off:
    \begin{equation}
    \Delta J = 4\left({ \epsilon_2 c_M\over d^2E/dJ^2} \right)^{1/2} = 4\left({ \epsilon_2 c_M\over d\omega/dJ}  \right)^{1/2} \ . \label{eq:width}
    \end{equation}    
We use this expression to produce the green regions in Figure \ref{2nd_overlap}(a),
    after inserting the exact $d\omega/dJ$ from Eq. (4.3.29) of \cite{1983_lichtenberg}.
% }\]

\subsection{Chaos from Secondary Resonance Overlap}\label{sec:overlap}
% Explain Creation of the Overlap Figure \[{
Chaos is postulated to occur where secondary resonances overlap with each other.
In Fig. \ref{2nd_overlap}(b), the overlapping region for the system in panel (a) is shaded red.
That red region is made as follows:
First, the green $[M, 1]$ secondary resonances from panel (a) are copied over,
    after inflating their width in $\Delta J$ by $\pi/2$ and changing their color to light green (to indicate their inflation); the reason for the inflation will be discussed shortly.
Second, the analysis that produced the light green region is repeated,
    but with the roles of the $\epsilon_1$ and $\epsilon_2$ resonances swapped.
That produces $[1, N]$ secondary resonances, in the notation introduced below Eq. (\ref{eq:phys}),
    which are shown in Fig. \ref{2nd_overlap}(b) in yellow, after inflation.\footnote{
    To be more precise, the yellow regions are made by first defining the swapped quantities $p_{\rm sw}$,
    $\phi_{\rm sw}$, etc., in which the 1 and 2 indices are swapped in their definitions (as described in \S \ref{sec:ikc}).
    The locations and widths are then calculated in the $p_{\rm sw}$-$\nu_{\rm sw}$ plane,
        and that result is transformed to the $p$-$\nu$ plane via the transformation in \S \ref{sec:ikc}. 
    Note that in calculating the $[1, N]$ resonances,
        we take the unperturbed Hamiltonian to be the first and third terms in Eq. (\ref{toy_model}),
        rather than Eq. (\ref{secpend}) with $M=0$; although the latter is more accurate,
        we choose the former for simplicity.
    \label{fn:swapped}}
The $[M, 1]$ resonances all lie to one side of the $[1, 1]$ resonance and the $[1, N]$ lie to the other.
Finally, the region where at least two of the secondary resonances overlap is colored red.
% }\]

% Explain Special Cases \[{ 
In the above procedure, three of the resonances must be treated specially: the $[1, 0]$, $[0, 1]$, and $[1, 1]$. 
For the $[1, 0]$, we do not include its overlap with the green regions in calculating the red.
That is because the $\epsilon_1$ resonance is incorporated into $H_{\rm pend}$ in Eq. (\ref{secpend}),
    and that equation is integrable for a single $M$, i.e., two $M$'s are needed for chaos.
We also neglect overlap with the $[0, 1]$ for the same reason.
For the third---the $[1, 1]$---its width may be calculated in two ways,
    as it is both an $[M,1]$ and a $[1,N]$ resonance.
We choose it to be the $[M,1]$ if the $\epsilon_1$ resonance is stronger ($\epsilon_1>\epsilon_2$); otherwise we choose it to be the $[1,N]$.
% }\]

% Reference back to the Improved Kick Criterion \[{
Also shown in Fig. \ref{2nd_overlap}(b) is the prediction of the improved kick criterion (black curve),
    which is copied from Fig. \ref{reverse_prediction}(c).
The two criteria (kick and overlap) are seen to largely agree, although the overlap criterion is spikier.
Comparing with the numerical integration (Fig. \ref{reverse_prediction}a),
    we see that most of the spikes in the numerical plot are attributable to overlapping secondary resonances. 
% }\]

% Compare SO to IKC \[{
Although the overlap criterion is more accurate than the kick criterion throughout most of the plot, 
    in the center of Fig. \ref{2nd_overlap}(c) overlap predicts less chaos;
    and in that region it is the kick criterion that is the more accurate,
    as may be seen by comparing with the numerical result.
That is because resonance overlap is only accurate when the widths of the two resonances are comparable, as discussed 
in \S \ref{sec:ooc} and \S \ref{sec:ikc}.
And near the center of the figure ($|\nu|\lesssim$ 2), neighboring resonances are ill-matched in size; e.g.,
    at the value of $\nu$ pointed to by the $[0,1]$ label in panel (a), the $[0,1]$ has a large width in $p$,
    while the $[1,1]$ is squeezed into the edges of the $\epsilon_1$ separatrix.
One could, of course, improve on the overlap prediction by applying the kick method to two secondary resonances,
    or considering the overlap of tertiary resonances \citep{1981_escande}.
But that lies beyond our scope.
% }\]

\subsection{Agreement With Improved Kick Criterion}\label{sec:agree}
% Derive Overlap \[{
\cite{1983_lichtenberg} demonstrate analytically that the overlap and kick criteria agree with each other,
    subject to the approximations of the classical kick criterion.
Here, we extend their analysis to the improved kick criterion.
From Eq. (\ref{eq:phys}), resonances $M$ and $M + 1$ differ in their frequency by $\delta\omega \approx {r\nu\over (M+lr)^2}$, when $M\gg 1$.
That translates into a spacing in actions of
    \begin{equation}
    \delta J = {r\nu\over (M+lr)^2}\left|{\td J\over \td\omega}\right|\ .
    \end{equation}
Overlap between neighboring secondary resonances occurs when $\Delta J>\delta J$ where $\Delta J$ is given in Eq. (\ref{eq:width}),\footnote{
    Strictly speaking, overlap occurs when the sum of the half-widths of adjacent resonances exceeds $\delta J$.
    We ignore this subtlety because adjacent resonances have comparable widths when $M \gg 1$.}
    i.e., when 
    \begin{align}
    16\epsilon_2 c_M > {(r\nu)^2\over (M+lr)^4}\left|{dJ\over d\omega}\right|  \ .\label{eq:marg}
    \end{align}
Setting ${\td J\over\td \omega} = {\td J \over \td E}{\td E \over \td T}{\td T \over \td\omega}$, ${dE\over dJ}=\omega$, ${dT\over d\omega}=-{2\pi\over\omega^2}$ and $\omega$ to its resonant value (Eq. \ref{eq:phys}) then yields
    \begin{align}
    \left|{dE\over dT}\right|&<{16\over 2\pi} {\epsilon_2}r\nu(M+lr)c_M\\
    &\approx {4\over \pi^2} {\epsilon_2}r^2\nu (\nu-\Delta p)A_{2r}[r(\nu-\Delta p)]
    \label{eq:dedt2}
    \end{align}
    where in the approximation we set $c_M$ to the form derived in Appendix \ref{sec:srd} (Eq. \ref{cm4}), and for simplicity we specialize to the upper circulation zone and set $\epsilon_1 = 1$.
Equation (\ref{eq:dedt2}) reproduces the improved kick criterion (Eqs. \ref{Ecrit0} \& \ref{jeremysDiscovery}), albeit with a numerical coefficient of ${4\over \pi^2}$ instead of unity \citep[as shown also in][]{1983_lichtenberg}.
Such a disagreement is not surprising, because both criteria are only defined up to arbitrary order-unity factors.
For this reason, we plot secondary resonances with widths inflated by $\pi/2$.
% }\]

\subsection{The 5:3--2:3 Crossing}\label{sec:apply}
% Compare to Chaos \[{
We return to the worked example in \S \ref{sec:we}, the 5:3--2:3 crossing,
    for which we repeat the procedure leading to  Fig. \ref{2nd_overlap}(b),
    and then transform to the $P_1$-$P_2$ plane with Eqs. (\ref{P1})--(\ref{P2}).
The result is shown in Fig. \ref{workednbody}(d),
    where now we depict all of the (inflated) secondary resonances light green.
The overlap region (red) is very similar to the numerical chaotic region from the perturbed pendulum model (panel b), which in turn agrees moderately well with what is found in N-body integrations (panel a).
The only place panels (b) and (d) disagree significantly is where there is a large disparity in resonance sizes,  which occurs in the region within the 5:3 that is also outside the 2:3.
A similar effect was seen in the previous example (Fig. \ref{2nd_overlap}).
% }\]

% Compare Numerical Integrations \[{
Much of the discrepancy between the two numerical results (panels a and b) may now be understood by comparing with panel (d). 
For example, the separatrix of the $[1, 1]$ is chaotic throughout panel (a), and its chaos is evidently due to its intersection with horizontal and vertical lines that represent high-order 2 body resonances.
Those resonances were excluded from the perturbed pendulum model. 
Many of the other secondaries in panel (a) are also chaotic, 
    due to their overlap with 2-body resonances. 
% }\]

\subsection{Three Body Resonances}
% Secondary Resonances are Three Bodies \[{
Secondary resonances in the perturbed pendulum model are three body resonances (3BR's) in the N-body problem. 
Here we discuss their locations and widths in the PP plane.
The 3BR resonant argument from an $[M, N]$ secondary is $M\phi + N\psi$,
    where for the previous example  $\phi = 5\lambda - 3\lambda_1$ and $\psi = 2\lambda - 3\lambda_2$, 
    and we neglect here the $\varpi$ terms.
The $[1, 1]$ resonance is the largest secondary (see below).
For the example, its argument is $\Phi_{[1, 1]} = 7\lambda - 3\lambda_1 - 3\lambda_2$.
Therefore, the location of the $[1, 1]$ is determined by ${d\over dt}\Phi_{[1, 1]} = 0\approx {7\over P}-{3\over P_1}-{3\over P_2}$ (continuing to neglect $\varpi$ terms), i.e., it traces out the following curve in the PP-plane:
    \begin{equation*}
    {P_2\over P} = {3\over 7-3P/P_1} \ ,
    \end{equation*}
    which matches what is seen in
    Fig. \ref{workednbody}(a) and (d).
% }\]

% Discuss Widths \[{
The width of an $\left[M, 1\right]$ resonance is worked out by \cite{1985_escande}.
We turn that into the 3BR width by inserting his Eq. (D.16) for $c_M$ into Eq. (\ref{eq:width}),
    and then converting $\Delta J$ to $\Delta P$,
    which yields
    \begin{align}
    \frac{\Delta P_{[M, 1]}}{P} &\approx \frac{\Delta P_1}{P}\sqrt{\frac{\epsilon_2}{\epsilon_1}}\p{\frac{\epsilon_1}{8E}}^{M/2}\Sigma\p{r, M} \label{Pwidth2b}
    \end{align}
    to lowest order in $\epsilon_1/E$, 
    where $E = p^2/2 - \epsilon_1\cos\phi_i$, and $\phi_i$ is the initial phase (see \S \ref{sec:we});
    and $\Sigma\p{r, M}$ is given by Eq. (D.17) of \cite{1985_escande} (with $\Sigma(r, M) = \Sigma_r^M$ in his notation).
Likewise, for the $\left[1, N\right]$ resonances, 
    \begin{align}
    \frac{\Delta P_{[1, N]}}{P} &\approx \frac{\Delta P_2}{P}\sqrt{\frac{\epsilon_1}{\epsilon_2}}\p{\frac{\epsilon_2}{8E_{\rm sw}}}^{N/2}\Sigma\p{\frac{1}{r}, N} \label{Pwidth1b}
    \end{align}
    where $E_{\rm sw} = p_{\rm sw}^2/2 - \epsilon_2\cos\psi_i$, in terms of the swapped quantities described in footnote \ref{fn:swapped}.
% }\]

% Discuss Scalings \[{
Of particular note is the dependence of the widths on $\epsilon_1$ and $\epsilon_2$,
    since those quantities are typically very small (of order $\mu e^k$; see Eq. \ref{eq:epsdef}).
One might naively have expected the $[1, 1]$ width to be $\sim \sqrt{\epsilon_1\epsilon_2}$,
    because it arises from the coupling of two MMR's,
    each with width $\sim \sqrt{\epsilon}$ (Eqs. \ref{dPin} and \ref{dPout}).
That would make the width extremely small.
But Eqs. (\ref{Pwidth2b}-\ref{Pwidth1b}) show that there is an extra factor of $E$ in the denominator,
    and that is typically of order $\epsilon_1$ in the region of interest.
Therefore the 3BR widths are not necessarily much smaller than those of the MMR's that generate them.
% ]\}

\subsection{Many Crossings}\label{sec:smc}
% Figure: PP Plot w/ Theory \[{
\begin{figure}[!tp]
    \includegraphics[width=1.0\linewidth]{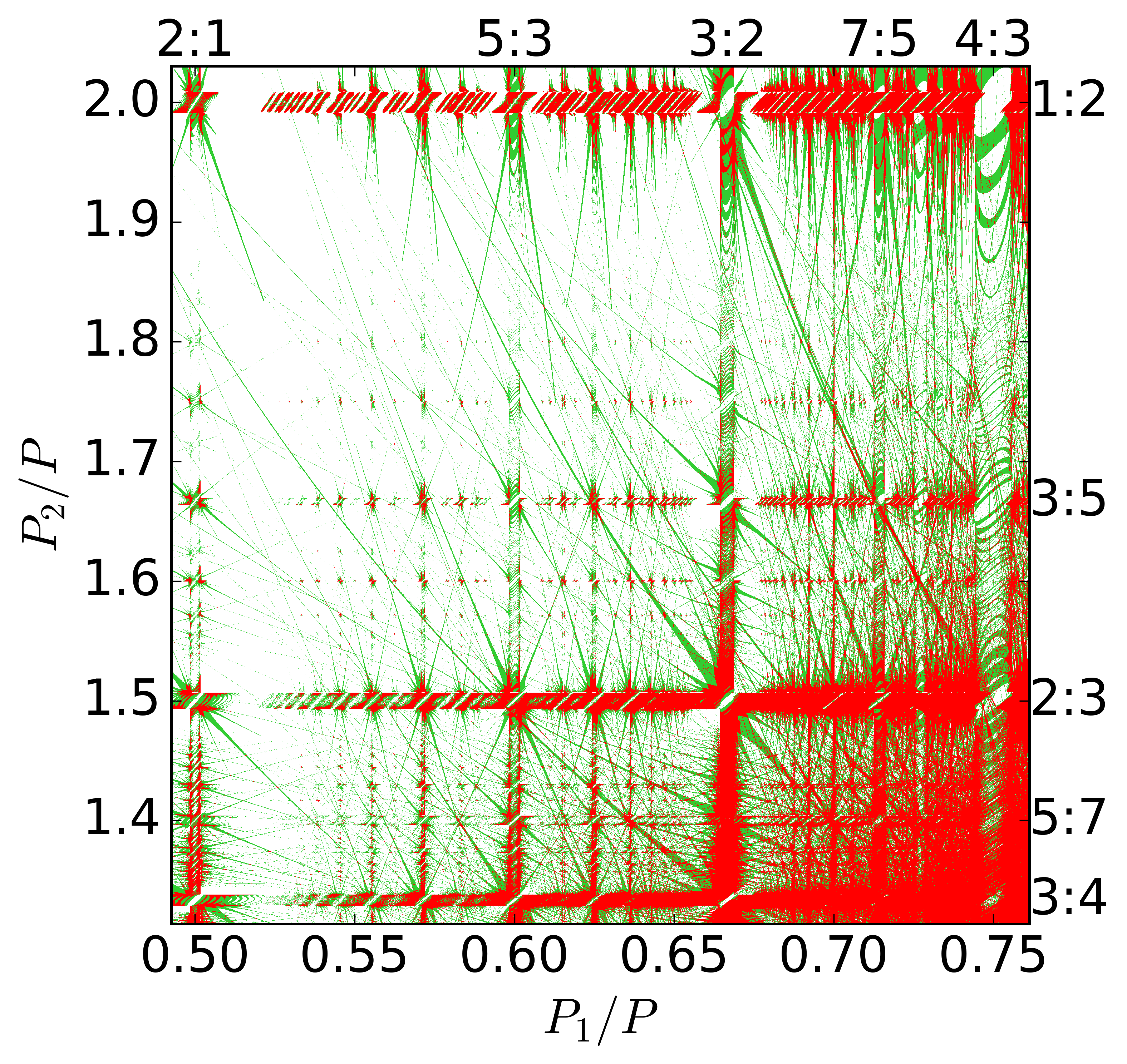}
    \caption{3BR's (green) and their overlap (red) for our fiducial system.
             We show all 3BR's that are at least three pixels in width.}
    \label{original_cross_2}
\end{figure}
% }\]

% Secondary Resonance Overlap Figure \[{
For the fiducial system of Fig. \ref{kepler-11},
the secondary resonances (i.e., 3BR's) are shown in
 Fig. \ref{original_cross_2} in green. And the regions where
 they overlap are shown in red. 
 Comparing to Fig. \ref{kepler-11}, we see that the strong chaos is adequately predicted,
    and some of the finer structure is captured.
Here, we consider also the overlap of secondary resonances from different two-body resonance crossings,
    which makes up a large portion of the chaos in the lower right of the figure.
Secondary resonances also interact with primary resonances to produce chaos,
   but we omit such crossings because they are subdominant.
% }\]

\begin{figure}[!tp]
    \includegraphics[width=1.0\linewidth]{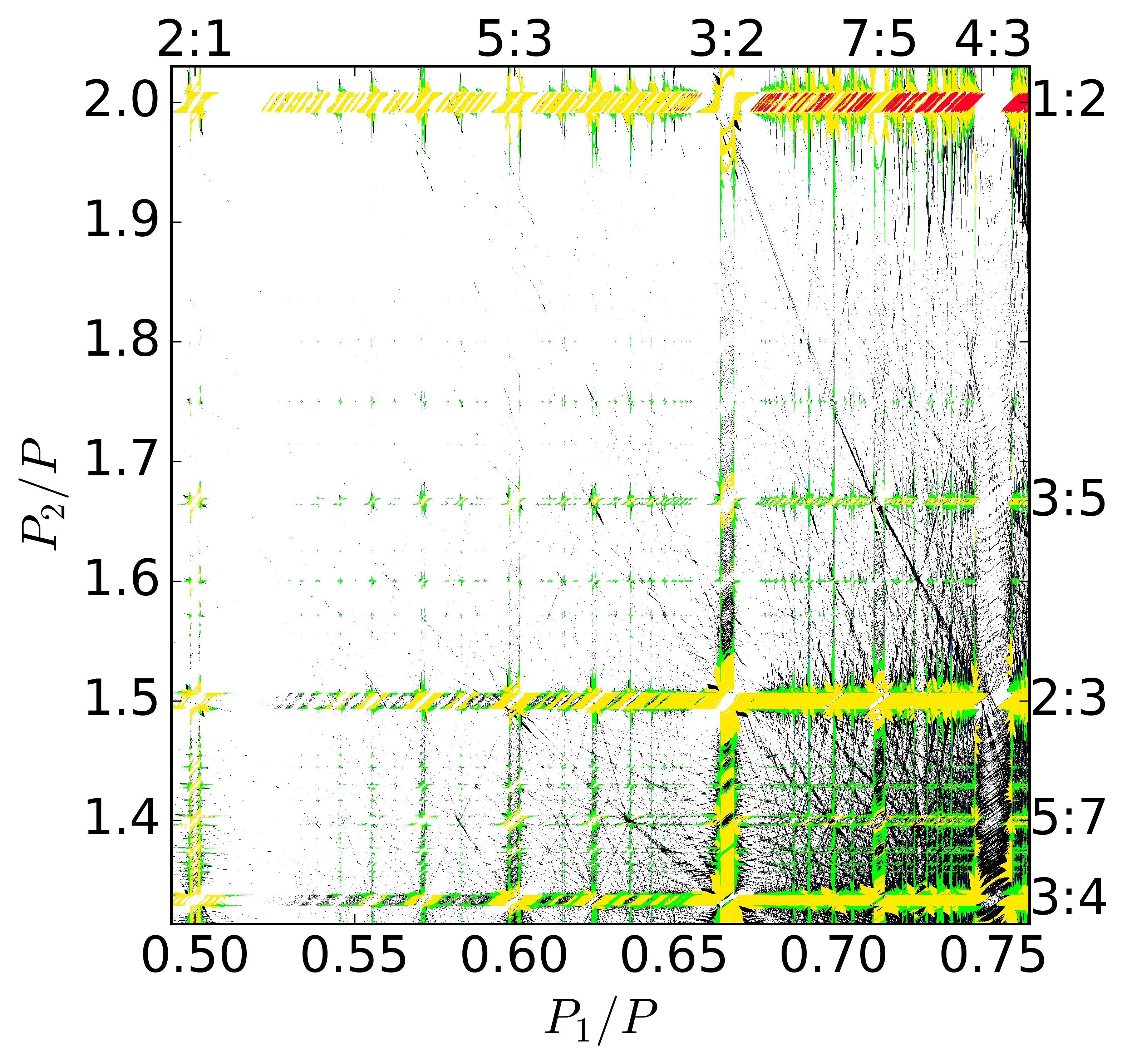}
    \vspace{-1 pt}
    \includegraphics[width=0.99\columnwidth]{"0-colorbar1"}
    \centering
    \caption{The predicted Lyapunov time based on overlapping secondary resonances.
    For two secondaries from the same 2BR crossing, 
    the prediction is color-coded; but for two 
    secondaries from different 2BR crossings, we are unable 
    to make a prediction, and those overlaps are colored black.}
    \label{original_cross_3}
\end{figure}

% Lyapunov Time \[{
We may also use secondary resonances to estimate Lyapunov times.
Empirically, we find that the Lyapunov time in a double surface of section is roughly constant throughout the $p$-$\nu$ plane.
Therefore, for each value of the parameters of the perturbed pendulum ($r,\epsilon_1,\epsilon_2$),
    we numerically evaluate the Lyapunov time at a given point in the $p$-$\nu$ plane;
    we choose the point where the two separatrices cross,
    which we find typically gives a good estimate for the Lyapunov time.
We then fit the Lyapunov time to a power-law expression, which results in
    \begin{align}
    t_{\rm lyap} \sim \frac{11}{\sqrt{r\epsilon_1}}\p{\frac{\epsilon_1}{\epsilon_2}}^{0.1} \label{lyap}
    \end{align}
In Fig. \ref{original_cross_3}, we show the Lyapunov times that result from the overlap of secondary resonances in Fig. \ref{original_cross_2}. 
For the case when the two secondaries come from the same 2BR crossing,
    we keep two different $M$ terms in Eq. (\ref{secpend}),
    and then rescale variables so that the Hamiltonian is that of the perturbed pendulum.
The Lyapunov time given by Eq. \ref{lyap} is indicated by the color in Fig. \ref{original_cross_3}.
For the case that the two secondaries come from different 2BR crossing,
    the calculation is more complicated.
Thus we merely color the resulting overlapping secondaries black.
There is order of magnitude agreement with Fig. \ref{kepler-11} in much of the plot.
The abundance of black in the lower-right corner highlights the importance of overlapping secondaries from different 2BR crossings.
% }\]

% Figure: PP Plot w/ Lyapunov Times \[{
% }\]

\section{Discussion}\label{sec:dis}
    
\subsection{Summary}\label{sec:recap}
Our principal results are as follows:
\begin{itemize}
 \item We mapped out the chaos in the PP plane, for a fiducial three-planet system where the middle planet is converted to a test particle (Fig. \ref{kepler-11}).
 As shown by Fig. \ref{kepler-11ro}, resonance overlap is at best a crude guide for where chaos occurs.
 
 \item To improve upon resonance overlap, we first reduced the dynamics near an MMR crossing to a much simpler system with 1.5 degrees of freedom: the perturbed pendulum (Eq. \ref{toy_model}).
 The reduction, which is based on a novel approximation that combines many subresonances (Appendix \ref{sec:rh}, \cite{2019_hadden}), 
 allows us to apply standard techniques from chaos theory to the planetary problem.
  The chaos map in the PP plane for a given MMR crossing is thereby reduced to a double surface of section of the perturbed pendulum (e.g., Figs. \ref{ni}a and \ref{workednbody}a-b).
 
 \item Two standard methods were applied to predict the chaotic zone. 
The first is kick theory (also known as the ``separatrix map'' or ``whisker map''),
    which predicts the chaotic zone around each resonance.
But that theory, as developed in the literature, is only adequate very near the separatrix.
We therefore developed an improved kick criterion that applies further away (\S \ref{sec:ikc} and Appendix \ref{app:ikc}).
A highly simplified prediction for the extent of the chaotic zones in the PP-plane is provided by Eqs. (\ref{PchaosI}) and (\ref{PchaosO}).
For our fiducial planetary system, we combined the results from many different MMR crossings
  in Fig. \ref{original_cross_1}, which agrees  with Fig. \ref{kepler-11} considerably better than naive resonance overlap. 
  
  \item The second standard method we applied is that of overlapping secondary resonances, or
  equivalently overlapping 3BR's (\S \ref{sec:so} and Appendix \ref{sec:srd}).
  We developed an improved theory for the overlap, which parallels the improved kick theory, and
  showed that both improved theories agree with each other.
  Moreover, overlapping 3BR's account for much of the finer structure in the PP plane.
  We also predicted the Lyapunov times (Eq. \ref{lyap} and Fig. \ref{original_cross_3}),
    which are in moderate agreement with those in Fig. \ref{kepler-11}. 
  \end{itemize}

\subsection{Validity of Key Assumptions}\label{sec:va}
We assess the validity of our key assumptions:  \begin{itemize}[listparindent=1.5em,parsep=-.5pt]
  \item Massless middle planet:
    Making the middle planet massive should not introduce  fundamentally new complications.
  For example, the reduction to a perturbed pendulum in Appendix \ref{sec:rh} proceeds virtually unchanged.
One complication is that whereas we only considered MMR crossings experienced by the middle planet, one must also consider crossings experienced by the inner and outer planets. 
  But if the three planets have comparable masses, those crossings should be subdominant---because the strongest interactions should involve the middle planet.
  As a check, we repeated Fig. \ref{kepler-11}, but with the middle planet having a mass of $12M_\oplus$ \citep{2011_lissauer}.
  We found very little difference from Fig. \ref{kepler-11}, beyond the slight broadening and strengthening of the chaotic regions.
  A second complication with a middle massive planet is its effect on the secular evolution, as discussed below.
    
 \item Three planets:  Based on the reasoning of the previous point, we  do not expect additional planets
 to introduce significant novelties. To determine whether any planet is chaotic, one need only 
 examine its MMR's with any two other planets.
 
  \item Coplanar planets:  Allowing for modest inclinations should typically have little effect,
        because inclinations are weakly coupled to eccentricities.
  As a test, we  repeated Fig. \ref{kepler-11} but with $i = e$ for each planet, and there was little difference to the figure.
  See also numerical experiments by \citet{Tamayo2021}.

  \item Pendulum model for MMR's (see footnote \ref{pendapprox}):
  the pendulum approximation breaks down for first-order MMR's at low eccentricity ($e\lesssim \mu^{1/3}\sim 0.03$, where the latter expression is for $10M_\oplus$ planets).
  In that case, the shape of the separatrix changes \citep{1983_Henrard},
  and one would have to modify the two methods (kick and secondary overlap) presented in this paper to account for this change.
  That lies beyond the scope of this work.

\item Neglect of secular effects: 
In adopting the pendulum approximation,
    we assume that the resonance strengths ($\epsilon_1$ and $\epsilon_2$) are constant,
    with values determined by the {\it initial} eccentricities.  
But secular evolution alters the eccentricities on a timescale of $\sim P/\mu$.
\citet{Tamayo2021} show that one may account for that effect in a simple way:
    rather than using the initial eccentricity,
    one should use the maximum eccentricity as determined by secular evolution alone. 
Note that our fiducial system exhibits little secular evolution---because the massless planet has higher $e$ than the other planets---and hence did not require such a correction.
But where such a correction is required, it is straightforward to apply.

An additional secular effect we neglect is chaos due to  overlapping  secular resonances.
That could potentially lead to chaos on even longer timescales, as in the Solar System.

\end{itemize}

\subsection{Comparison with Prior Work on Overlapping 3BR's}\label{sec:compare}
% Comparison to Petit \[{
\cite{Petit2020}, building on \cite{2011_quillen} and \cite{2014_quillen},
    determine a criterion for chaos in three planet systems from the overlap of 3BR's.
In contrast to us, they assume the planets' orbits are circular,
    which allows for much more closely spaced planets before chaos occurs.
For example, for our fiducial example of Fig. \ref{kepler-11},
    they predict that the threshold for chaos for equal-spaced planets occurs at $P_1\approx 0.88P$ and $P_2\approx 1.14P$,
    which is beyond the range of Fig. \ref{kepler-11}.
% }\]

% Details of 3 Bodies \[{
Nonetheless, it is instructive to compare their 3BR's with ours.
Their 3BR locations are the same as those of the 3BR's that arise from the 1:1--1:1 crossing.
The angles of the two MMR's are $\phi=\lambda-\lambda_1$ and $\psi=\lambda-\lambda_2$,
    and so the crossing produces 3BR's with angle $(M+N)\lambda-M\lambda_1-N\lambda_2$.
In the PP plane, these 3BR's describe the curves $M{P\over P_1}+N{P\over P_2}=M+N$,
    all of which intersect the point $(P_1, P_2) = (1, 1)P$.
Therefore those authors omit many of the 3BR's that we consider, as is appropriate at zero eccentricity.
To determine the widths of the aforementioned 3BR's, \cite{2011_quillen} considers the combination of
two MMR's with angles $I_1(\lambda-\lambda_1)$ and $I_2(\lambda-\lambda_2)$, respectively, for integers $I_1$ and $I_2$.
The subsequent papers by \cite{2014_quillen} and \cite{Petit2020} incorporated
an additional set of crossings into the width calculation: those from two 1st order MMR's, at integer
combination $[M,N] = [1,1]$, which is easily seen to produce the same 3BR arguments as the 1:1--1:1 crossings.
% }\]

\subsection{The Outer Solar System}\label{sec:oss}
% Figure: Solar System \[{
\begin{figure}
    \includegraphics[width=0.99\columnwidth]{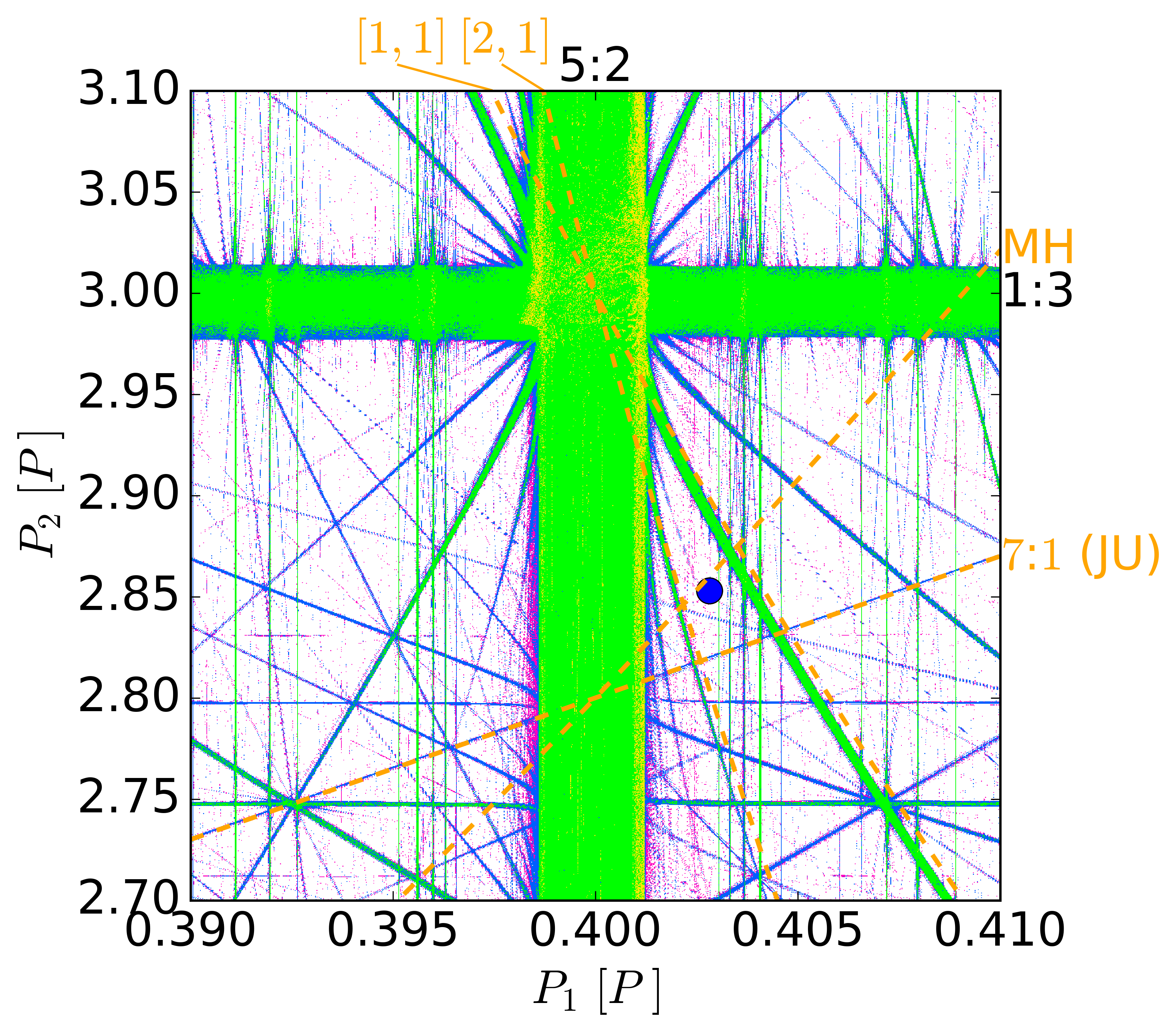}
    \vspace{-1 pt}
    \includegraphics[width=0.99\columnwidth]{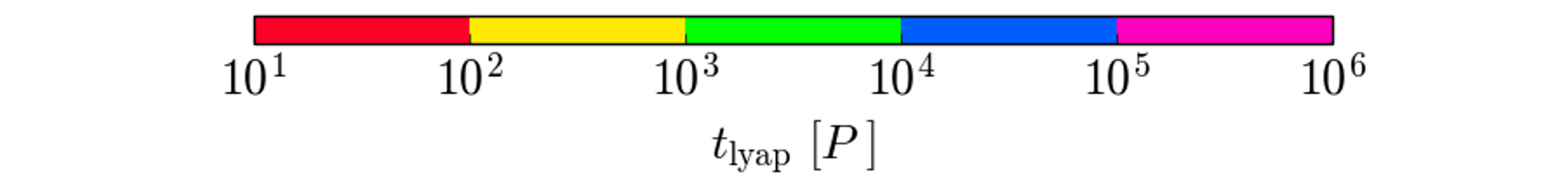}
    \centering
    \caption{PP map for the Jupiter-Saturn-Uranus system.
             The parameters are from JPL's DE441 ephemerides \citep{2021_park} evaluated at Jan. 1, 2000.
             The blue dot indicates the true system.
             The orange dashed lines denote the nominal locations of various resonances:
                the one labeled MH is the 3BR proposed by \cite{1999_murray_holman};
                the ones labeled $[1,1]$ and $[1,2]$ are the corresponding $[M,N]$ combinations from the 5:2-1:3 crossing;
                and the one labeled 7:1 is a resonance between Jupiter and Uranus.
            The dashed lines do not correct for secular frequencies,
                which would shift the lines by a small amount ($\sim 0.001P$).}
    \label{solar_system}
\end{figure}
% }\]

% Solar System Chaos \[{
Chaos in the outer Solar System has been attributed to 3BR's amongst Jupiter, Saturn, and Uranus \citep{1999_murray_holman}.
In Fig. \ref{solar_system}, we show the PP map for those three planets.  
We see that the strongest chaos in the vicinity of the true system is dominated by the crossing between the 5:2(Jup.-Sat.)--1:3 (Sat.-Ura.), as well as the 3BR's that result from the $[1,1]$ and $[2,1]$ combinations of those two resonances.
Nonetheless, the current system is close to---not inside of---those chaotic zones.
Instead, its chaos  appears to be more strongly affected by a very high order resonance between Jupiter and Saturn, as well as by a secular resonance.
We defer a more careful analysis to future work. 

\cite{Guzzo2005} analyzes the Solar System in a manner very similar to Fig. \ref{solar_system},
    although with more zoomed-in axes, in which the  1:3 is not evident.
We have also repeated Fig. \ref{solar_system} with Neptune present, as did \cite{Guzzo2005}.
Similar to him, we find that Neptune has a minor effect on the overall chaos,
    beyond introducing some extra weak chaos.
Finally, we note that the specific 3BR identified by \cite{1999_murray_holman} is the $[1, 1]$ combination that arises from the crossing of the 5:2(Jup.-Sat.) and the 1:7(Jup.-Ura.).
And although there is some chaos associated with the 1:7, there does not appear to be any chaos associated with the combination proposed by \cite{1999_murray_holman}, even when we repeat Fig. \ref{solar_system} at a significantly higher resolution.
% }\]

\acknowledgments
This research was supported in part through the computational resources and staff contributions provided for the Quest high performance computing facility at Northwestern University which is jointly supported by the Office of the Provost, the Office for Research, and Northwestern University Information Technology.
Y.L. acknowledges NASA grant NNX14AD21G and NSF grant AST-1352369.
S.H. acknowledges support from the CfA Fellowship and support by the Natural Sciences and Engineering Research Council of Canada (NSERC), funding reference \#CITA 490888-16.
\software{Matplotlib \citep{Matplotlib}, NumPy \citep{numpy}, Pandas \citep{pandas}, SciPy \citep{scipy}, REBOUND \citep{2012_rein}, WHFast \citep{2015_rein}, SPOCK \citep{spock}.}

\clearpage
\appendix

\section{Reduction to Perturbed Pendulum}\label{sec:rh}
\subsection{Circular Planets}\label{sec:circ}
% State Energy in Circular Case \[{
We consider a test particle perturbed by two resonances: a $j_1$:$j_1 - k_1$ resonance with an inner planet (denoted with subscript 1) and a $j_2$:$j_2 + k_2$ resonance with an outer planet (denoted with subscript 2), where the $j_i$ and $k_i$ are all positive.
We derive here the test particle's  Hamiltonian when the two perturbing planets are on circular orbits
(see 
 \S \ref{sec:ecce} for the extension to eccentric planets).
The Hamiltonian (or, really, energy) is
    \begin{align} H = -\underbrace{\frac{GM}{2a}}_{H_\t{Kep}} - \frac{GM}{a}\p{\underbrace{\mu_1C_1e^{k_1}\cos\pd{j_1\lambda - \pd{j_1 - k_1}\lambda_1 - k_1\varpi}}_{H_{\t{res}, 1}} + \underbrace{\mu_2C_2e^{k_2}\cos\pd{j_2\lambda - \pd{j_2 + k_2}\lambda_2 + k_2\varpi}}_{H_{\t{res}, 2}}} \label{H_tp_only}\end{align}
    where $\set{a, e, \lambda, \varpi}$, without subscripts, are the standard orbital elements for the test particle; subscripted orbital elements are for the planets; $M$ is the stellar mass; 
    $\mu_i$ is the planet mass scaled to the stellar mass; 
    and the $C_i$ are disturbing function coefficients.\footnote{
    A more accurate model may be obtained by replacing the coefficients $C_ie^{k_i}$ in Eq. \eqref{H_tp_only}, which approximate the resonance amplitudes at leading order in eccentricity, with the cosine amplitudes $S_{j_i, k_i}(\alpha,e)$ defined in Eq. (23) of \cite{2018_hadden}, which are correct to all orders in eccentricity, then proceeding through
    the derivation presented here. Nonetheless, we find that the $C_ie^{k_i}$ provide sufficiently good approximations of the resonance amplitudes for all cases considered in this paper so we have chosen
    to work with these expressions as they are more common in the literature.}
$C_1$ is obtained directly from Eq. (6.113) of \cite{1999_murray_dermott},
    while $C_2$ has an additional factor of $a/a_2$.
For $k_i\leq 4$, \cite{1999_murray_dermott} provide a convenient table  in their Appendix B:
    for example, for a second order resonance with the inner planet ($k_1 = 2$), $C_1 \rightarrow f_{53}$ in their table's notation, and for a second order with the outer planet ($k_2=2$), $C_2 \rightarrow \frac{a}{a_2}f_{45}$.
% }\]

% Perturbed Pendulum Derivation - Enumerated \[{
The derivation of Eq. (\ref{toy_model}) proceeds as follows:
\begin{enumerate}
% Reduce H_{res} \[{
\item
We adopt the pendulum approximation, in which we treat all quantities in the non-Keplerian part of Eq. (\ref{H_tp_only}) as constants, except for $\lambda$ and $\lambda_i$.
The validity of this approximation is examined in \S \ref{sec:va}.
Setting $\lambda_i = n_it$ for the two planets, where $n_i$ is their mean motion, we have
    \begin{align} H_{\rm res, 1} + H_{\rm res, 2} = \epsilon_1\cos\pd{j_1\pd{\lambda - n_{r, 1}t}} + \epsilon_2\cos\pd{j_2\pd{\lambda - n_{r, 2}t}} \end{align}
    where
    \begin{eqnarray}
    \epsilon_1 = \mu_1|C_1|e^{k_1} \ , \ \ 
    \epsilon_2 = \mu_2|C_2|e^{k_2} \ , 
    \label{eq:epsdef}\\
    n_{r, 1}= {j_1 - k_1\over j_1}n_1  \ , \ \ 
    n_{r, 2}= {j_2 + k_2\over j_2}n_2 \ .
    \end{eqnarray}
Here, $n_{r, i}$ is the mean motion when the test particle is at nominal resonance with the $i$th planet.
We drop constant phases within the cosine arguments because they have little influence on the dynamics (although they are needed for transforming between orbital and pendulum co-ordinates).\footnote{
    Without neglecting phases, the argument of each cosine term is $j_i\pd{\lambda - n_{ri}} - \xi_i$ where $\xi_1 = k_1\varpi_1 - \pd{j_1 - k_1}\lambda_{1, {\rm initial}}$ and $\xi_2 =  - k_2\varpi_2 + \pd{j_2 + k_2}\lambda_{2, {\rm initial}}$.
    If $C_i < 0$, then $\xi_i \rightarrow \xi_i + \pi$. \label{foot:phase}}
% }\]

% Deal with Keplerian Part \[{
\item
We turn Eq. (\ref{H_tp_only}) into a proper Hamiltonian by replacing $a$ in $H_{\rm kep}$ with $\Lambda\equiv\sqrt{GMa}$, which is the momentum conjugate to $\lambda$.
We then expand $\Lambda$ around its value at nominal resonance with either planet 1 or 2.
Typically, one wishes to expand around whichever resonance has the larger effect.
For definiteness, we choose planet 1 and comment on how things change with the other choice in \S \ref{sec:rpp}.
Therefore, we write
    \begin{align} H_{\rm kep} = -{G^2M^2\over 2\Lambda^2} \approx n_{r, 1}\Lambda_{r, 1}\p{{\Lambda - \Lambda_{r, 1}\over\Lambda_{r, 1}}-{3\over 2}\p{{\Lambda-\Lambda_{r, 1}\over\Lambda_{r, 1}}}^2} \label{hkep}\end{align}
    where $\Lambda_{r, 1}=\sqrt{GMa_{r, 1}}$ and $a_{\rm r, 1} = (GM/n_{\rm r, 1}^2)^{1/3}$. 
% }\]

% Canonical Transformation \[{
\item
We make a canonical transformation to the new co-ordinate and momentum, $\set{\phi,P}=\set{j_1(\lambda-n_{r1}t),(\Lambda-\Lambda_{r1})/j_1}$,
    which produces a new Hamiltonian which differs from the original only in that the first term in brackets in Eq. (\ref{hkep}) disappears.
Then, in order to remove a relative constant factor between $H_{\rm kep}$ and $H_{\rm res,i}$, we rescale both the new momentum and Hamiltonian by the same constant, ($-\Lambda_{r1}/\sqrt{3}j_1$);
    although that rescaling is not canonical, it leaves the equations of motion unchanged.
The result is
    \begin{eqnarray}
    \label{eq:full_hamiltonian_model}
    H(\phi,p,t) = \p{n_{r, 1} j_1\sqrt{3}}\times \p{{p^2 \over 2} - \epsilon_1\cos\phi - \epsilon_2\cos\p{{j_2\over j_1}\pd{\phi - \pd{{n_{r1} - n_{r2}}}j_1 t}}} \label{hpphi}
    \end{eqnarray}
where $p$ is the rescaled momentum (given explicitly below).
% }\]

% Rescale Time and Finish \[{
\item
In order to remove the overall constant in front of Eq. (\ref{hpphi}), we rescale time: $t_{\rm rescale}=t(n_{r, 1}j_1\sqrt{3})$.
We therefore have for our final Hamiltonian:
    \begin{eqnarray} H(\phi, p, t_{\rm rescale}) = {p^2\over 2} - \epsilon_1\cos\phi - \epsilon_2\cos\pd{r\pd{\phi - \nu t_{\rm rescale}}} \label{eq:hfin}\end{eqnarray}
    where
    \begin{eqnarray}
    r&=& {j_2\over j_1} \label{eq:rdef} \\
    \nu &=& {1\over \sqrt{3}}{n_{r2} - n_{r1}\over n_{r1}}
    \label{eq:nudef} \ ,
    \end{eqnarray}
    and $\phi$ and $p$ are related to the orbital elements and (unscaled) time via
    \begin{eqnarray}
    \phi&=& j_1(\lambda - {n_{r1}t}) \label{eq:defphi} \\
    p&=&-\sqrt{3} {\Lambda - \Lambda_{r1}\over \Lambda_{r1}}\approx - {\sqrt{3}\over 2}{a - a_{r1}\over a_{r1}}
    \label{eq:deflast} \ .
    \end{eqnarray}
% }\]
\end{enumerate}
% }\]

\subsection{Eccentric Planets}\label{sec:ecce}
When the two planets are eccentric, each $H_{\rm res, i}$ in Eq. (\ref{H_tp_only}) becomes a sum of $k_i$ cosine terms, e.g.
    $H_{\rm res,1}=\mu_1\left( C_{1,0} e^{k_1}\cos(\psi - k_1\varpi) + C_{1,1} e_1e^{k_1 - 1}\cos(\psi - \varpi_1 - (k_1 - 1)\varpi) + \cdots\right)$ where $\psi = j_1\lambda - (j_1 - k_1)\lambda_1$.
If one had to consider each of those cosine terms separately, the calculations needed for this paper would be exceedingly cumbersome.
But one may avoid that with a trick \citep[][hereafter H19]{2019_hadden}.
Following H19, we first write for the inner planet's resonance
    \begin{eqnarray} H_{{\rm res},1}=\mu_1{\rm Real} \left\{\left[\sum_{l=0}^{k_1} C_{1,l}\left(e_1{\rm e}^{-i\varpi_1}\right)^{l}\left(e{\rm e}^{-i\varpi}\right)^{k_1-l}\right]{\rm e}^{i\psi}\right\} \end{eqnarray}
    where the Roman ${\rm e}$ is an exponential (Euler's number).
H19 showed that the square-bracketed sum in the above equation may be approximated by a single combined term.
Moreover, whereas the sum apparently depends on the complex eccentricities ($e{\rm e}^{-i\varpi}$) of the test particle and inner planet separately,
    the combined term only depends on a single combined quantity.
To be explicit, H19 shows that the square-bracketed term is sufficiently well approximated by
    \begin{eqnarray}
    [...] \approx {C}_{1}\left(\tilde{e}_1{\rm e}^{-i\widetilde{\varpi}_1}\right)^{k_1} \ ,
    \label{eq:cnew}
    \end{eqnarray}
    where the new quantities (with tilde's) are defined via
    \begin{eqnarray}
    \tilde{e}_1e^{-i\widetilde{\varpi}_1} = ee^{-i\varpi} - \p{\frac{a_1}{a}}^{0.825}e_1e^{-i\varpi_1} \ ,
    \label{ediff}
    \end{eqnarray}
    which is nearly the difference between the two complex eccentricities.
We note parenthetically that H19 obtain the coefficient on the second term in the above expression by inserting Eq. (\ref{ediff}), with an undetermined coefficient, into Eq. (\ref{eq:cnew}), with an undetermined amplitude\footnote{
    While H19 fits for both undetermined parameters,
    we have found that $C_1$ is a sufficiently good approximation for the amplitude and use it instead.
    This is true in all circumstances except when indirect terms are present.
    In our main example (Fig. \ref{kepler-11}), this is only significant for the 2:1 and 1:2 resonances,
    so we fit two parameters in those cases.};
    then expanding into $k_1+1$ terms; and finally choosing the undetermined coefficients by matching term-by-term with the square-bracketed sum.
Remarkably, even though there are only two fitting parameters and $k_1 + 1$ coefficients to match, the error after the fit is performed is small.
Some discussion of why this fit works can be found in \S 2.4 of \cite{2019_hadden}.
Nonetheless, we may take advantage of this result by using for $H_{\rm res, 1}$ the term displayed in Eq. (\ref{H_tp_only}), after simply replacing $e\rightarrow \tilde{e}_1$ and $\varpi\rightarrow \widetilde{\varpi}_1$.
The resonance with the outer planet proceeds in a nearly identical way:
    in $H_{\rm res,2}$, one makes the replacement $e\rightarrow \tilde{e}_2$ and $\varpi\rightarrow \widetilde{\varpi}_2$,
    where the variables with tildes are defined via $\tilde{e}_2{\rm e}^{-i\widetilde{\varpi}_2} = e{\rm e}^{-i\varpi} - (a_2/a)^{0.825}e_2{\rm e}^{-i\varpi_2}$.

The Hamiltonian with eccentric planets is therefore identical to the one with circular planets derived above (Eqs. \ref{eq:hfin}--\ref{eq:deflast}),
    except that the eccentricity that enters in $\epsilon_1$ should be $\tilde{e}_1$ (rather than $e$), and the one that enters in $\epsilon_2$ should be $\tilde{e}_2$.

\section{The Melnikov-Arnold (MA) Integral}\label{sec:MA}
\setcounter{equation}{0}

\begin{figure}[!h]
    \begin{center}
        \includegraphics[width=0.38\textwidth]{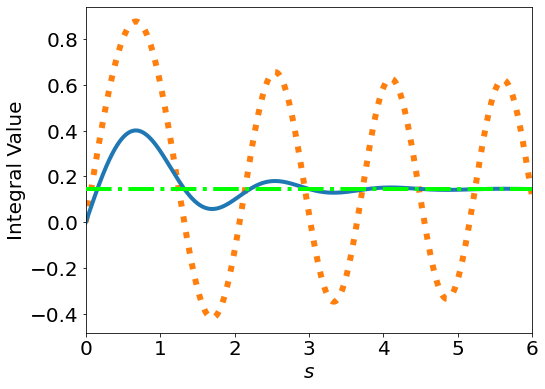}
    \end{center}
    \caption{An illustration of proper integral versions of $A_{2r}(r\nu)$ (orange dotted line) and $\tilde{A}_{2r}(r\nu)$ (blue solid line) as functions of the integration bound, as well as the analytic expression given in Eq. (\ref{MA2}) (horizontal green line), with $r = 1$ and $\nu = 4.2$.
    This figure is based on Fig. 3.22 in \cite{1983_lichtenberg}.}
    \label{MA_illustration}
\end{figure}

% Define MA Integral \[{
The MA integral is\footnote{We set $\epsilon_1=1$ throughout this appendix.}
    \begin{equation}
    A_{2r}(r\nu) \equiv \lim_{s\rightarrow \infty}\int_{-s}^{s}\cos\left[ r\left(\phi_{\rm sep}(t')-\nu t'\right)  \right]dt' \label{MAdef}
    \end{equation}
    where $\phi_{\rm sep}$ is the pendulum's phase on the separatrix \citep{1979_chirikov}.
Technically, the integral does not converge in the limit of large $s$, but oscillates around a constant value as $s$ increases.
That constant value has the following analytic expression:
    \begin{align}
    \left[A_{2r}(r\nu)\right]_{\rm constant} = \frac{2\pi}{\Gamma(2r)}\frac{e^{\pi r\nu/2}}{\sinh\pd{\pi r\nu}}\pd{2 r\nu}^{2r - 1}\pd{1 + f_{2r}(r\nu)} \label{MA2}
    \end{align}
    where $f_{2r}$ is given by Eq. (A.9) of \cite{1979_chirikov};
    in particular, $f_{2r}=0$ when $0\leq r\leq 1$, and $f_{2r}$ is typically order unity for $1<r\lesssim 2$.
The dependence of $A_{2r}$ on $s$ (before the $s$ limit is taken) is illustrated by the orange dotted curve in Fig. \ref{MA_illustration}.
At large $s$, it oscillates around Eq. (\ref{MA2}) (the horizontal line).
Even though Eq. (\ref{MAdef}) is not strictly identical to Eq. (\ref{MA2}), we  follow \cite{1979_chirikov},
    and drop the ``constant'' label from Eq. (\ref{MA2}) throughout the body of the paper.
% }\]

% Figure: MA Integral Illustration \[{
% }\]

% Alternate MA Integral \[{
The integral that appears in the classical kick criterion (\S \ref{sec:ckc}) is not Eq. (\ref{MAdef}), but
    \begin{align}
    \tilde{A}_{2r}(r\nu) &\equiv \frac{1}{\nu}\lim_{s\rightarrow\infty}\int_{-s}^{s}\dot{\phi}_{\rm sep}\pd{t'}\cos\left[ r\left(\phi_{\rm sep}(t')-\nu t'\right)\right]~\td t'\ , \label{MAnew}
    \end{align}
    which has an extra factor of $\dot{\phi}_{\rm sep}/\nu$.
Unlike Eq. (\ref{MAdef}), this integral does converge to Eq. (\ref{MA2}).
The blue curve in Fig. \ref{MA_illustration} shows this integral and its convergence at large $s$.
We note that in the context of the classical kick criterion, the issue of integral non-convergence is merely an artifact of the standard definition of the MA integral (Eq. \ref{MAdef});
    i.e., had one chosen to adopt Eq. (\ref{MAnew}) as the definition, no such issue would arise.
We have belabored the non-convergence because it plays an important role
in the improved kick criterion  (Appendix \ref{app:ikc}).
% }\]

\section{Improved Kick Criterion}\label{app:ikc}
\setcounter{equation}{0}

\begin{figure*}[!h]
    \includegraphics[width=0.98\textwidth]{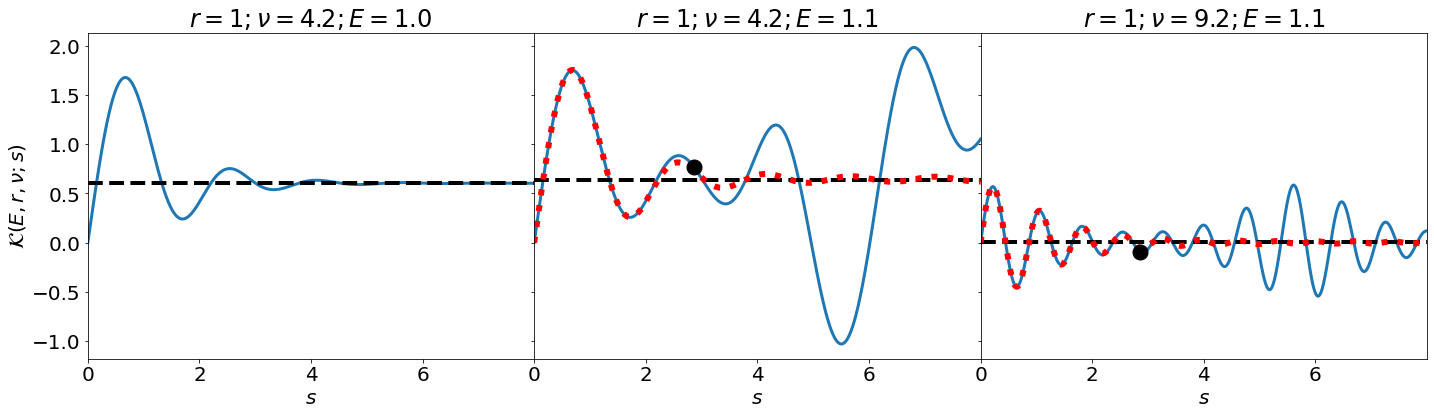}
    \centering
    \caption{In each panel, we show the original definition of ${\cal K}$ as a function of $s$ (Eq. \ref{K_true}) as a blue solid line.
             We show our approximation to ${\cal K}$ (Eq. \ref{still_dp}) as a red dotted line.
             The approximation is absent from the left panel because it is identical to Eq. (\ref{K_true}) when $E=1$.
             The horizontal dashed lines are given by Eq. (\ref{K_mine}).
             The black points indicate the value of Eq. (\ref{K_true}) at $s = T / 2$.
             It is absent in the left panel because $T = \infty$ there.}
    \label{fig:oscillations}
\end{figure*}

% Explain Problem with K(E, r, \nu) \[{
For the kick criterion, one must evaluate ${\cal K}$ in Eq. (\ref{calK}),
    which we restate here for convenience as
    \begin{eqnarray}
    {\cal K}(E, r, \nu;s) = \int_{-s}^{s}\dot{\phi}_{\rm unp}(t')\cos\left[r\pd{\phi_{\rm unp}(t') - \nu t'}\right]\td t' \label{K_true}
    \end{eqnarray}
    where $s = {T/2}$ and $T = T(E)$ is the unperturbed period.
Note that we now allow ${\cal K}$ to depend explicitly on the integral limit $s$,
    for reasons that will be apparent;
    we set $\epsilon_1 = 1$ for this appendix;
    and we explicitly display the dependence of ${\cal K}$ on $r$ and $\nu$,
    unlike in the body of the paper.
For the classical criterion, one sets $\phi_{\rm unp}$ in this expression to be on the separatrix and $T=\infty$.
But for the improved criterion, we allow $\phi_{\rm unp}$ to follow a trajectory slightly displaced from the separatrix by approximating
    \begin{align}
        \phi_{\rm unp}(t)\approx \phi_{\rm sep}(t) + \pd{\Delta p} t \label{new_phi}
    \end{align}
    where $\Delta p$ is a constant that we take to be the relative drift rate at $t = 0$:
    \begin{align}
        \Delta p &= \dot{\phi}_{\rm unp}\pd{0} - \dot{\phi}_{\rm sep}\pd{0}\\
                 &= \sqrt{2\pd{E + 1}} - 2\ . \label{eq:dp}
    \end{align}
Inserting this $\phi_{\rm unp}$ into Eq. (\ref{K_true}) leads to
    \begin{equation}
    {\cal K}\pd{E, r, \nu; s} \approx \int_{-s}^{s} \p{\dot{\phi}_\t{sep}\pd{t'} + \Delta p}\cos\pd{r\pd{\phi_\t{sep}\pd{t'} - \pd{\nu - \Delta p} t'}} \td t' \ . \label{still_dp}
    \end{equation}
Figure \ref{fig:oscillations} shows ${\cal K}$ versus $s$ for three sets of parameters.
The left panel has $E = 1$ (and hence $\Delta p = 0$) and is the same as the blue curve in Fig. \ref{MA_illustration}.
In the middle panel, the trajectory is displaced from the separatrix ($E = 1.1$).
The solid blue curve shows the exact integral (Eq. \ref{K_true}); the red dotted curve shows the approximation (Eq. \ref{still_dp}); and the black dot indicates where $s=T/2$. 
We see that the approximation agrees well for this value of $E$ at the time of interest. 
The right panel shows another comparison at higher $\nu$.
% }\]

% Figure: K vs s \[{
% }\]

% Description of E vs t Figure \[{
Fig. \ref{fig:oscillations} also illustrates  an important, yet subtle, complication: the value of Eq. (\ref{K_true})  (i.e., the height of the black dot at time $T/2$) is subject to fast oscillations,
    and it depends on the precise phase of the fast oscillation at that time.
Yet in truth, one wishes to average out the fast oscillations.
That is because each successive kick is subject to similar fast oscillations, and so they average out after many kicks.
As a result, the value of Eq. (\ref{K_true}) at time $T/2$ is not what should enter the kick criterion.
One requires the height of the horizontal lines in the figure, rather than that of the black dots.
% }\]

% Calculate K \[{
To calculate the horizontal lines, we start from Eq. (\ref{still_dp}), and focus first on the $\dot{\phi}_{\rm sep}$ term within the brackets.
At large $|t|$, $\dot{\phi}_{\rm sep}$ decays smoothly to zero ($\dot{\phi}_{\rm sep}\approx 4e^{-|t|}$). Therefore, extending the integration limits  to $s=\infty$ will remove oscillations on a timescale much faster that $T$.
The second bracketed term ($\Delta p$) is subdominant to the first for $|t'|<T/2$ and is oscillatory as $s\rightarrow\infty$.
Therefore, we neglect it.
The net result is that Eq. (\ref{K_true}) should be replaced with
    \begin{align}
    {\cal K}\pd{E, r, \nu} &\approx \int_{-\infty}^{\infty} \dot{\phi}_\t{sep}\pd{t'}\cos\pd{r\pd{\phi_\t{sep}\pd{t'} - \pd{\nu - \Delta p} t'}} \td t'\\
        &= \pd{\nu - \Delta p}A_{2r}\left[r\pd{\nu - \Delta p}\right] \label{K_mine}
    \end{align}
    where the latter equality follows from Appendix \ref{sec:MA}, and $A_{2r}$ is given by Eq. (\ref{MA2}).
Equation (\ref{K_mine}) is the main result of this appendix;
    the horizontal lines in Fig. \ref{fig:oscillations} are at the values prediction by this equation.
    In \S \ref{sec:ikc}, we use Eq. (\ref{K_mine}) to produce the improved kick criterion.
% }\]

% Explain Answer Physically \[{
The ``improved'' expression for ${\cal K}$ differs from the classical one (Eq. \ref{K_chi}) by the replacement $\nu\rightarrow \nu-\Delta p$.
One may understand why $\nu-\Delta p$ is the relevant frequency by referring to a surface of section,
    such as Fig. \ref{sos_pict}.
The perturbing resonance is at height $p = \nu$,
    and our trajectory of interest is at height $p=2+\Delta p$ at the critical moment when it receives its kick (i.e., when $\phi_{\rm unp}\approx 0$).
The difference between those two heights is equal to the difference in frequencies, which sets the strength of the kick.
Therefore the kick strength on our trajectory of interest is almost equivalent to the kick on a trajectory {\it on} the separatrix,
    provided the height of the perturbing resonance is lowered from $\nu\rightarrow \nu-\Delta p$.
% }\]

\begin{figure*}[!h]
    \includegraphics[width=0.9\textwidth]{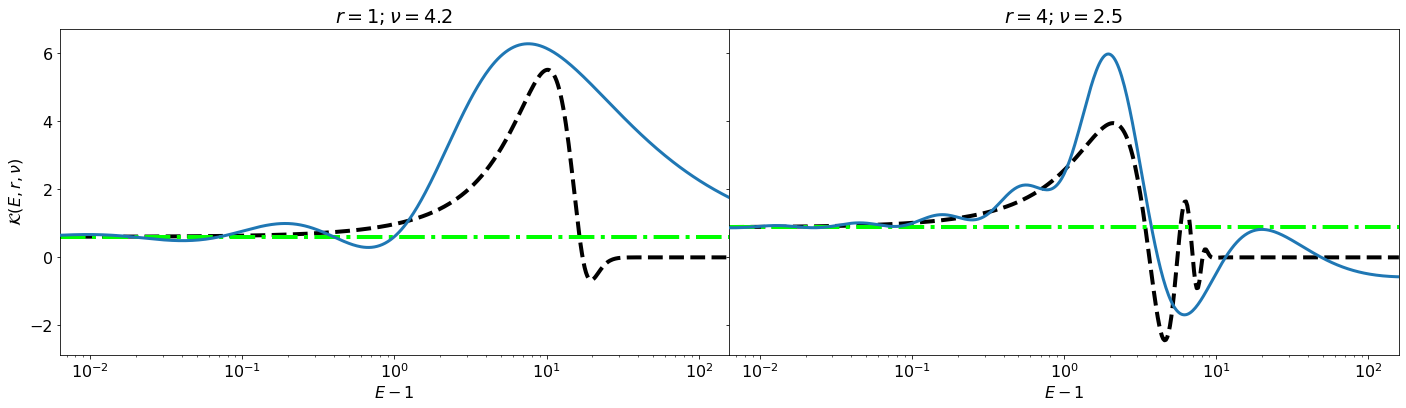}
    \centering
    \caption{A comparison of different expressions for ${\cal K}$ as a function of E: 
             Solid blue lines depict the original definition (Eq. \ref{K_true}),
             green dot-dashed lines depict Chirikov's approximation (Eq. \ref{K_chi}),
             and black dashed lines depict our approximation (Eq. \ref{K_mine}).
             The left panel has the same parameters as the two left panels of Fig. \ref{fig:oscillations}.}
    \label{fig:dE}
\end{figure*}

% Introduce Figure \[{
Fig. (\ref{fig:dE}) (left panel) compares the various expressions for ${\cal K}$ as a function of $E$, for $r=1$ and $\nu=4.2$.
The horizontal dot-dashed line is the classical ${\cal K}$ (Eq. \ref{K_chi}), which is independent of $E$.
The blue solid curve is the exact expression before the removal of fast oscillations (i.e., Eq. \ref{K_true} at $s=T/2$).
One sees that very close to the separatrix ($E-1\lesssim 1$), the exact expression oscillates around the classical one.
The oscillations are caused by the aforementioned fast oscillations, and do not contribute to chaos.
In fact, the classical expression is more accurate than the exact one at such close distances to the separatrix.
But the peak in the blue curve at $E-1\sim 10$ is not due to fast oscillations,
    and the classical expression is inadequate there.
The black dashed curve is the improved expression (Eq. \ref{K_mine}).
It both captures the true behavior close to the separatrix,
    and does an adequate job at capturing the peak---considerably better than the classical criterion.
The right panel of Fig. \ref{fig:dE} is similar, but for $r=4$ and $\nu=2.5$, and likewise shows how Eq. (\ref{K_mine}) captures the broad peak while removing the fast oscillations.
We note that a truer measure for how well Eq. (\ref{K_mine}) performs can be ascertained by comparing its prediction for the transition to chaos with numerical integrations, as is done in Fig. \ref{reverse_prediction} (a) and (b) for the case $r=1$ and $\epsilon_2/\epsilon_1=0.8$.
We have done so for many additional cases, and find that Eq. (\ref{K_mine}) invariably produces the best match with simulations.
% \]}

% Figure: calK vs E \[{
% }\]

% K from Secondary Resonances \[{
We conclude this Appendix with a second, more formal, derivation of Eq. (\ref{K_mine}), based on the action-angle formalism derived in Appendix \ref{sec:srd}.
The rate of change of the unperturbed energy is
    \begin{equation}
    {dE\over dt}={\partial E\over \partial J}{dJ\over dt}=-\omega{\partial H\over \partial\theta}
     = \epsilon\omega\sum_{M = -\infty}^{\infty}c_M\pd{M + lr}\sin\left[M\theta + r\pd{l\theta - \nu t}+\psi^{(1)} \right]
    \end{equation}    
    where $\omega = 2\pi/T$; the final expression follows from the Fourier
    expansion of the perturbation in the angle variable (Eq. \ref{ftd});
    and $\psi^{(1)}$ is the forcing phase at the time of the first kick (\S \ref{sec:km}).
Integrating over an (unperturbed) orbit yields a change in energy with the form of Eq. (\ref{eq:dek}) where
    \begin{equation}
    {\cal K}(E,r,\nu)=\sum_{M=-\infty}^\infty
    \omega c_M\pd{\frac{M + lr}{r}}\int_{-\frac{\pi}{\omega}}^{\frac{\pi}{\omega}} \cos\left[M\theta + r\pd{l\theta - \nu t}\right]\td t
    \end{equation}
We keep in this sum over $M$ only the dominating term (i.e., the one that is near-resonant) satisfying $\pd{lr + M}\omega - r\nu \sim 0$.
This effectively removes the fast oscillations.
Next, we set $\theta = \omega t$ and evaluate the integral to yield
    \begin{eqnarray}
    {\cal K}\pd{E, r, \nu} \approx 2\omega c_M\p{\frac{M + lr}{r}}\frac{\sin\left[\frac{\pi}{\omega}\pd{M\omega + r\pd{l\omega - \nu}}\right]}{\pd{M + lr}\omega - r\nu}\ .
    \end{eqnarray}
Using the approximation that $\pd{lr + M}\omega - r\nu \sim 0$ and inserting Eq. (\ref{cm4}) for $c_M$, we get
    \begin{align}
    {\cal K}\pd{E, r, \nu} &\approx \pd{\nu - \Delta p}A_{2r}\left[r\pd{\nu - \Delta p}\right] \label{K_mine2}
    \end{align}
    which is identical to Eq. (\ref{K_mine}).
% }\]

\section{Perturbed Pendulum Expansion in Action-Angle Variables}\label{sec:srd}
\setcounter{equation}{0}

% Deal with Fourier Expansion \[{
For \S \ref{sec:so}, we require the expansion of the perturbed piece of the Hamiltonian, $H'$, in terms of the action-angle variables of the unperturbed pendulum.
Before expanding $H' \equiv \epsilon_2\cos[r(\phi-\nu t)]$, we first expand $e^{ir\phi}$.
In the libration zone of $H_{\rm pend}$, $\phi$ is a periodic function of $\theta$ (i.e. $\phi\pd{\theta} = \phi\pd{\theta + 2\pi n}$ where $n \in\Z$).
Hence we may expand
    \begin{align} e^{ir\phi}=\sum_M c_M e^{iM\theta} \ , {\rm where\ } c_M\equiv {1\over 2\pi}\int_{-\pi}^\pi e^{i(r\phi-M\theta)}\td\theta {\ \rm (in\ libration\ zone)} \label{lib} \end{align}
But in the circulation zone, $\phi$ changes by $2\pi$ when $\theta$ does; i.e., it is the difference $(\phi-\theta)$ which is a periodic function of $\theta$.
In that case, one should replace $\phi \rightarrow \phi - \theta$ throughout Eq. (\ref{lib}). 
Both zones may be combined by introducing 
    \begin{align} l = \begin{cases} 0 {\rm\ in\ libration\ zone} \\ 1 {\rm\  in\ circulation\ zones} \end{cases} \end{align}
    in which case
    \begin{align} e^{ir\phi} &= \sum_M c_M e^{i(M + lr)\theta}\ ,\ {\rm where\ } c_M \equiv {1\over 2\pi}\int_{-\pi}^\pi e^{i\left[r\phi-(M + lr)\theta\right]}\td\theta.\label{cmapp} \end{align}
Note that $c_M$ is real, which follows from the fact that $H_{\rm pend}$ is unchanged when $\phi\rightarrow -\phi$.
Using Eq. (\ref{cmapp}), we write
    \begin{align} H' &= \epsilon_2\cos\left[r\p{\phi - \nu t}\right] \label{btd}\\ &= \epsilon_2\sum_{M = -\infty}^{\infty}c_M\cos\left[M\theta + r\p{l\theta - \nu t}\right] \label{ftd}\end{align}
    where
    \begin{align} c_M &= \frac{1}{2\pi}\int_{-\pi}^{\pi}\cos\left[r\phi - \p{M + lr}\theta\right]\td\theta\label{cm}\ . \end{align}
This expression was derived for the libration zone by \cite{1978_smith} and for the circulation zone by \cite{1981_escande}.
The above expressions for $H'$ and $c_M$ are exact, and are the main result of this appendix.
There are several approaches to approximate/compute the integral in $c_M$:
\cite{1978_smith} give analytic expressions for $r = \frac{1}{2}$ and $r = 1$, and show analytic solutions exist for all $2r\in\Z$.
\cite{1985_escande} gives approximate expressions far from the separatrix for both inside and outside of resonance, which we use for the sake of computation time to create Figs. \ref{original_cross_2}, and \ref{original_cross_3}.
% }\]

% Use for Appendix C \[{
We conclude this appendix with an approximate expression for $c_M$, which is needed for deriving the improved kick criterion with the action-angle method (Appendix \ref{app:ikc}) and for quantifying secondary resonance overlap (\S \ref{sec:overlap}).
As was done to evaluate Eq. (\ref{K_true}), we set $\phi\pd{t} \approx \phi_s\pd{t} + \pd{\Delta p} t$ and take the
integration bounds to infinity to get
    \begin{align}
    c_M &\approx \frac{\omega}{2\pi}\int_{-\infty}^{\infty}\cos\left[r\phi_{\rm sep} - \p{M + lr}\omega t + r\Delta p~t\right]\td t\label{cm2}\\
    &= \frac{\omega}{2\pi}A_{2r}\left[\p{M + lr}\omega - r\Delta p\right] \label{cm3}\end{align}
    where $A_{m}\pd{\lambda}$ is the MA-integral given in Eq. (\ref{MA2}).
Although the above approximation is generally a good one, it fails at $M=0$.
In that case, as $E\rightarrow \infty$ the argument of the $A_{2r}$ approaches a constant (because $\omega\rightarrow \Delta p+2$) and the prefactor ($\omega$) increases without bound.
However, in truth $\lim_{E\rightarrow\infty}c_0 = 1$. To correct for that, we change the prefactor by a quantity that is small whenever $M\ne 0$, yet yields the correct limiting behavior of $c_0$, by setting
     \begin{align}
     c_M &\approx \frac{1}{2\pi}\left[\omega - \frac{r\Delta p}{M + lr}\right]A_{2r}\left[\p{M + lr}\omega - r\Delta p\right]\ . \label{cm4}
     \end{align}
% }\]

% Add Bibliography \[{
\bibliography{paper}{}
\bibliographystyle{aasjournal}
% }\]
\end{document}